\documentclass[article,oneauthor,pdftex,10pt,a4paper]{mdpi-arxiv}

\Title{Quantum-classical entropy analysis for nonlinearly-coupled continuous-variable
bipartite systems}

\Author{\'Angel S. Sanz$^{\ddagger}$*\orcidA{}}

\AuthorNames{A. S. Sanz}

\address{%
Department of Optics, Faculty of Physical Sciences,
Universidad Complutense de Madrid,\\
Pza.\ Ciencias 1, Ciudad Universitaria, 28040 Madrid, Spain;\\
$^{\dagger}$ a.s.sanz@fis.ucm.es}

\corres{Correspondence: a.s.sanz@fis.ucm.es; Tel.: +34-91-394-4673}

\abstract{The correspondence principle plays a fundamental role in quantum mechanics, which
naturally leads us to inquire whether it is possible to find or determine close classical
analogs of quantum states in phase space---a common meeting point to both classical and
quantum density statistical descriptors.
Here, this issue is tackled by investigating the behavior of classical analogs arising
upon the removal of all interference traits displayed by the Wigner distribution functions associated with a given pure quantum state.
Accordingly, the dynamical evolution of the linear and von Neumann entropies is numerically computed for a continuous-variable bipartite system, and compared with the corresponding classical counterparts, in the case of two quartic oscillators nonlinearly coupled under regular and chaos conditions.
Three quantum states for the full system are considered: a Gaussian state, a cat state,
and a Bell-type state.
By comparing the quantum and classical entropy values, and particularly their trends,
it is shown that, instead of entanglement production, such entropies rather provide us
with information on the system (either quantum or classical) delocalization.
This gradual loss of information translates into an increase in both the quantum and the classical realms,
directly connected to the increase in the correlations between both parties' degrees of freedom which, in
the quantum case, is commonly related to the production of entanglement.
}


\newcommand{\bd}{\begin{displaymath}}
\newcommand{\ed}{\end{displaymath}}
\newcommand{\be}{\begin{equation}}
\newcommand{\ee}{\end{equation}}
\newcommand{\ba}{\begin{eqnarray}}
\newcommand{\ea}{\end{eqnarray}}

\usepackage{xcolor}
\usepackage{soul}

\begin{document}


\section{Introduction}
\label{sec1}

Entanglement can be regarded as the characteristic trait of quantum mechanics, recalling
Schr\"odinger \cite{schrodinger:ProcCamPS:1935}, which means it plays a central role, as
an essential resource, in the implementation and development of modern quantum technologies
\cite{bouwmeester-bk:2000,nielsen-chuang-bk}.
The capability to transfer information at long distances between two entangled parties without
a physical interaction mediating between them \cite{schrodinger:ProcCamPS:1935,EPR:PhysRev:1935}
sets a crucial difference with respect to two classically correlated systems, where
information transmission necessarily requires the active action of physical interactions.
This is consistent with the common statement that there is no classical analog for entanglement.
However, this seems in contradiction with the fact that, by virtue of the correspondence
principle, classical mechanics should approximate quantum mechanics in one way or another, or,
at least, manifest in some form.
Thus, despite the uniqueness of entanglement as a genuine quantum trait, some relationships
between the classically chaotic dynamics exhibited by nonlinear systems and entanglement have 
been found and reported in the literature.
For instance, it has been found that the amount of entanglement increases in rates proportional
to the corresponding Lyapunov exponents
\cite{zurek:PhysicaD:1995,furuya:PRL:1998,sarkar:PRE:1999,ghose:PRA:2004,ghose:PRE:2004}.
Indeed, this trend has also been considered the other way around, i.e., using the increase
of entanglement as a signature of quantum chaos \cite{zurek:PhysicaD:1995,ghose:PRE:2004}.

To further inquire about the issue, the common formal ground rendered by a Liouvillian or
phase-space formulation, common to both classical and quantum mechanics, seems to be appropriate.
In the quantum case, this is possible by describing the system state and its evolution within
the Weyl--Wigner--Moyal representation
\cite{weyl:ZPhys:1927,wigner:PhysRev:1932,moyal:MathProcCambrdPhilSoc:1949}.
In this representation, the density matrix $\hat{\rho}$ accounting for the system state in an
$N$-dimensional configuration space is recast in terms of the corresponding Wigner quasi-distribution
function \cite{ballentine-bk}:
\begin{equation}
 \rho_W({\bf r}, {\bf p}) = \frac{1}{(\pi\hbar)^N}
 \int \langle {\bf r} - {\bf s}| \hat{\rho} | {\bf r} + {\bf s} \rangle
   e^{2i{\bf p} \cdot {\bf s}/\hbar} d{\bf s} ,
 \label{wigner}
\end{equation}
where $({\bf r}, {\bf p})$ is a $2N$ phase-space vector, with ${\bf r} = (r_1, r_2, \ldots , r_N)$
and ${\bf p} = (p_1, p_2, \ldots , p_N)$ being, respectively, the position and momentum vectors in
such phase space.
An advantage of this nonlinear representation is that it allows us to pass from a $\mathbb{C}^N$
complex space to a $\mathbb{R}^{2N}$ real space to describe the quantum system.
This is, however, at the expense of translating interference traits, the distinctive signature of
quantum coherence, into negative-definite regions in the associated phase space.
The time evolution for $\rho_W$ is described by the Moyal equation
\be
 \frac{\partial \rho_W({\bf r}, {\bf p}, t)}{\partial t} =
  - \{ \{ \rho_W({\bf r}, {\bf p}, t), H({\bf r}, {\bf p}, t) \} \}
 = - \{ \rho_W({\bf r}, {\bf p}, t), H({\bf r}, {\bf p}, t) \}_P + {\rm HOC}(\hbar^2) ,
 \label{eq1}
\ee
where $\{ \{ \cdot , \cdot \}\}$ is the Moyal bracket, defined as
\be
 \{ \{ \cdot , \cdot \}\}
 \equiv \frac{2}{\hbar} \sin \left[ \frac{\hbar}{2}
 \left( \overleftarrow{\nabla}_{\bf r} \overrightarrow{\nabla}_{\bf p}
 - \overleftarrow{\nabla}_{\bf p}  \overrightarrow{\nabla}_{\bf r} \right) \right] ,
 \label{moyal}
\ee
and $\{ \cdot , \cdot \}_P$ is the usual Poisson bracket, which arises from the lowest order of
the Taylor expansion of the Moyal bracket, Equation~(\ref{moyal}).
The term ${\rm HOC}(\hbar^2)$ gathers all higher-order contributions of such Taylor
expansion, which only includes explicit even powers of $\hbar$ (other than a possible implicit
dependence on this parameter in the Wigner distribution function).

Formally, quantum--classical correspondence relies on the presence of the above HOC term.
To better appreciate this statement, consider that the (phase space) Hamiltonian has the
usual form $H({\bf r},{\bf p}) = {\bf p}^2/2m + V({\bf r})$.
In this case, the HOC term reads as
\be
 \mathcal{O}(\hbar^2) = \sum_{n \ge 1} \left(\frac{\hbar}{2}\right)^{2n} \frac{(-1)^n}{(2n+1)!}
  \ \nabla_{\bf r}^{2n+1} V({\bf r}) \ \nabla_{\bf p}^{2n+1} \rho_W ({\bf r}, {\bf p}, t) .
\ee
Accordingly, the quantum dynamics will resemble classical dynamical behaviors
whenever this term becomes meaningless, e.g., whenever $\hbar \to 0$, or if the potential is a
polynomial of degree two or smaller.
This is the reason why coherent (Glauber) states, for instance, can be regarded as classical
states.
If, on the other hand, the spatial variations of the potential function $V({\bf r})$ are relevant
(higher than the second order), as it is the case of nonlinear potentials, particularly those
inducing classically chaotic dynamics, the behavior of the Wigner distribution function will
quickly develop
typical quantum features, e.g., extensive negative-definite regions associated with the
appearance of interference traits.
However, this does not prevent the quantum system to be somehow influenced by an underlying classical
dynamics mediated by the first term in the r.h.s.\ of Equation~(\ref{eq1}).
This would explain, for instance, the above-mentioned relationship between quantum dynamics and
(classical) Lyapunov exponents.

Nonetheless, there is also the possibility to observe important quantum effects with a vanishing
HOC term.
Think, for instance, of a simple superposition of two Gaussian wave packets.
In this case, the direct analogy between Equation~(\ref{eq1}) and its classical counterpart breaks down
immediately, even though the evolution equation is exactly the same in both cases.
This is only as, in the quantum case, we allow the phase-space distribution (or quasi-distribution,
strictly speaking) to be negative-definite in certain space regions, while classically we do not,
which, on the other hand, is necessary in order to preserve the unitarity of ${\rm Tr}[\hat{\rho}^2]$
and hence the purity of the quantum state.
Now, note that only in the case of (classical) Gaussian density distributions that preserve their
Gaussianity all the way through the time evolution (i.e., in those cases where the potential
function is a polynomial of degree two or lesser), the unitarity is preserved.
This does not mean that a quantum and a classical Gaussian distribution are equivalent.
Due to coherence preservation, the former is able to exhibit interference
if it is superimposed with another Gaussian.
This is not possible, though, in the classical case, because the corresponding distribution
($\rho_{\rm cl}$) makes reference to an incoherent swarm of particles, i.e., particles that
have nothing to do one with the others.
However, some works in the literature have dealt with the issue of determining classical-type
eigendistributions in the Liouville phase space, analogous to
the Wigner eigendistributions in the case of the quantum harmonic oscillator
\cite{brumer-jaffe:JPC:1984,brumer-jaffe:JCP:1985,brumer-jaffe:PRL:1985}
and also in the case of the evolution of anharmonic oscillators \cite{milburn:PRA:1986}.

Now, in general, ${\rm Tr}[\rho_{\rm cl}^2] \le 1$, which considers the case of
any arbitrary classical distribution, and follows from the Liouville theorem.
The issue has been investigated earlier on in the context of quantum--classical
correspondence for decoherence induced by nonlinear couplings in bipartite systems
\cite{brumer-gong:PRL:2003,brumer-gong:PRA:2003,brumer-gong:JModOpt:2003}.
The elementary tool to explore this correspondence is the computation of entropies associated
with one of the two subsystems, such as the linear and the von Neumann entropies.
These entropies can be understood~\cite{nielsen-chuang-bk} as a measurement of entanglement
produced over time by quantifying the degree of mixedness of one of the parties.
Note that, as it was pointed out by Schr\"odinger \cite{schrodinger:ProcCamPS:1935}, as soon
as two systems start interacting, their state can no longer be described in a separate manner, regardless of whether they are spatially separated (typical EPR condition
\cite{EPR:PhysRev:1935}) or not.
An increase in the amount of entanglement means an increase in our lack of knowledge on the
system state, which, in turn, translates into an increase in the corresponding entropy measure.
This issue presents a remarkable similarity with former studies relating chaos with energy
relaxation rates
\cite{brumer-hamilton:PRA:1982,brumer-hamilton:JCP:1983,brumer-christoffel:PRA:1986}
or decoherence
\cite{brumer-pattnayak:PRL:1996,brumer-pattnayak:PRL:1997,brumer-pattnayak:PRE:1997,shepelyansky:PRA:2003}.
Furthermore, it is significant that entanglement dynamics of initially Gaussian states
can be fully described by means of a classical entropy in the classical limit \cite{furuya:PRA:2005}.

In order to evaluate how close classical distributions may approximate quantum ones in entropy
measures, here we reexamine this problem considering two nonlinearly coupled one-dimensional
systems, such that depending on the coupling strength the corresponding classical dynamics can
be either regular or chaotic.
In particular, we shall focus on the analysis of the time evolution exhibited by the entropy
of one of the subsystems.
This behavior will be analyzed not only in the case of Gaussian states, for which the initial
classical counterpart is clear, but also with (superposition) cat states and Bell-type
entangled states, which are non-classical states with interest both in quantum mechanics
as well as in quantum optics
\cite{dodonov:Physica:1974,dodonov-manko-bk:2003,sanders:PRA:1992,sanders:JMO:1993,sanders:JPhysA:2012}.
In these cases, the classical analogs will be determined from the corresponding Wigner
distribution functions,
removing the negative-definite parts.
This choice relies on the fact that, in the classical limit, the rapidly oscillatory behavior
displayed by the Wigner distribution function generates negative-definite regions with negligible area.
Indeed, as it is shown, in the classical limit, both the quantum system and its classical
counterpart display analogous entropy rates.
Accordingly, we provide an interpretation for the increase in the quantum entropy, which makes
clear not only why it approaches the classical one, but also why it increases in the classical limit.
This might seem counter-intuitive, as the linear and von Neumann entropies are
supposed to be quantifiers of entanglement~\mbox{\cite{plenio-vedral:PRL:1997,plenio-vedral:PRA:1998,horodecki:QuantumInfComp:2001,bruss:JMathPhys:2002}}, which, on the other hand, has no classical counterpart.
This behavior thus leads to an alternative manner to interpret these measures, as quantifiers of
the maximum delocalization undergone by the system, a well-defined concept in both quantum and
classical mechanics.
Accordingly, due to the incoherence nature of the swarms of trajectories associated with
classical distributions, it is seen that the classical entropies set an upper bound over the
corresponding quantum entropies.
The fact that the latter remain lower can be associated with a higher information compression
in quantum mechanics, which is possible due to the coherence swapping between the two entangled~parties.

The work is organized as follows.
The specific functional form of the linear and von Neumann entropies in phase space is introduced
in Section~\ref{sec2} for classically and quantum mechanically.
Moreover, to be self-contained, some general properties are also discussed.
The working model and corresponding numerical simulations carried out are presented and
discussed in Section~\ref{sec3}.
The definitions of the classical analogs to the quantum states considered are also introduced
in this section.
These continuous-variable bipartite states include Gaussian (classical) states, cat states, and
Bell-type entangled states.
The general picture arising from the results described in Section~\ref{sec3} is provided in
Section~\ref{sec4}.
To conclude, a series of summarizing remarks are given in Section~\ref{sec5}.


\section{Quantifying Entanglement in Phase Space}
\label{sec2}

Different measurements have been proposed in the literature to quantify entanglement~\mbox{\cite{plenio-vedral:PRL:1997,plenio-vedral:PRA:1998,horodecki:QuantumInfComp:2001,bruss:JMathPhys:2002}}.
Among these entanglement quantifiers, the most commonly used ones for pure states are the linear
and the von Neumann entropies
\cite{wehrl:RMP:1978,wehrl:RepMathPhys:1979,wehrl:RepMathPhys:1991,popescu-rohrlich:PRA:1997}.
For mixed states, other more general quantifiers have been proposed, such as the
entanglement of formation \cite{bennett-smolin:PRA:1996}, the entanglement of distillation
\cite{bennett-smolin:PRL:1996,verstraete:PRA:2003}, or the negativity
\cite{peres:PRL:1996,horodecki:PhysLettA:1996,duan-cirac-zoller:PRL:2000,simon:PRL:2000}.
Here, given that only pure states are considered to describe the full or joint system,
the use of linear and von Neumann entropies suffices to carry out the quantum--classical analysis.
Furthermore, because the joint state remains pure along time, the amount of entanglement of both
parties is the same at any time (which can easily be proven using the Schmidt decomposition),
thus ensuring that it can be measured with respect to any of the two subsystems.

Thus, let us assume that $X$ and $Y$ represent two one-dimensional continuous-variable systems,
specified by their positions $x$ and $y$, respectively.
Without any loss of generality, the subsystem $X$ will be taken as the reference system,
and $Y$ as the environment, which will be traced out.
The reduced density matrix describing the state of $X$ is
$\tilde{\rho}_X = {\rm Tr}_Y ( \hat{\rho} )$, where $\hat{\rho}$ is the density matrix
accounting for the $XY$ joint system.
Accordingly, the linear entropy is defined as
\begin{equation}
 S_L \equiv 1 - {\rm Tr}_X ( \tilde{\rho}_X^2 ) .
 \label{eq3}
\end{equation}
This is a direct measure of the loss of purity---determined by the quantity
${\rm Tr}_X ( \tilde{\rho}_X^2 )$---undergone by $X$ as it becomes more entangled with $Y$
due to the interaction, i.e., as its degree of mixedness (classicality) increases.
For a pure state, given that its purity is 1, the linear entropy vanishes.
Note that the idempotence property of the density matrix is preserved in this case, so
$\tilde{\rho}_X^2 = \tilde{\rho}_X$ and, consequently,
${\rm Tr}_X ( \tilde{\rho}_X^2 ) = {\rm Tr}_X ( \tilde{\rho}_X ) = 1$.
For a mixed state, however, this property gets lost, and
${\rm Tr}_X ( \tilde{\rho}_X^2 ) < {\rm Tr}_X ( \tilde{\rho}_X ) = 1$.
Thus, in sum, we have
\begin{equation}
 0 \le S_L \le 1 ,
 \label{eq45}
\end{equation}
where the upper bound is approached as the degree of mixedness increases.
An evident advantage of this quantity is its relative simplicity to provide us with
information about the amount of entanglement between the two parties by observing how one
of them gradually losses its purity (i.e., as the action of decoherence becomes more
relevant).
This has been beneficially used, for instance, in studies relating chaotic dynamics with
production of entanglement \cite{furuya:PRL:1998,sarkar:PRE:1999,ghose:PRA:2004}.

In the Wigner representation, Equation~(\ref{eq3}) reads as
\begin{equation}
 S_L = 1 - 2\pi\hbar \int \tilde{\rho}_W^2(x,p_x) dx dp_x ,
 \label{eq47}
\end{equation}
where
\be
 \tilde{\rho}_W(x,p_x) = {\rm Tr}_Y \left[ \rho_W (x,y,p_x,p_y) \right]
  = \int \rho_W (x,y,p_x,p_y) dy dp_y .
\ee
If the Wigner distribution function provides us with a phase-space picture of the quantum system,
analogously in classical statistical mechanics we have density distributions accounting for the
statistical distribution of phase-space trajectories.
Therefore, we can introduce the classical analog of Equation~(\ref{eq47}) as
\begin{equation}
 S_L^{\rm cl} = 1 - 2\pi\hbar \int \tilde{\rho}_{\rm cl}^2(x,p_x) dx dp_x ,
 \label{eq46}
\end{equation}
where $\tilde{\rho}_{\rm cl}(x,p_x)$ specifies a classical density distribution in the reduced
phase subspace associated with the reference system $X$, obtained after integrating over all
other phase-space coordinates (those related to $Y$, in this case).

Regarding the von Neumann entropy, it is a generalization of the Shannon entropy used in
classical information theory \cite{shannon:BellSystTechJ:1948-1,shannon:BellSystTechJ:1948-2},
and is defined as
\begin{equation}
 S_V \equiv - {\rm Tr}_X [\tilde{\rho}_X \ln \tilde{\rho}_X]
 = - \sum_i \lambda_i \log \lambda_i ,
 \label{eq2}
\end{equation}
where $\lambda_i$ denotes the eigenvalues of the reduced density matrix.
This quantity determines the amount of information that is stored in the quantum state describing
the system of reference, which increases with its degree of mixture, and is considered to be the
best measure of entanglement for pure states \cite{popescu-rohrlich:PRA:1997}.
Apart from its connection to thermodynamics~\cite{popescu-rohrlich:PRA:1997}, this entropy
measure has also some properties that make it of interest in quantum statistical mechanics
and quantum information theory
\cite{nielsen-chuang-bk,wehrl:RMP:1978,wehrl:RepMathPhys:1979,wehrl:RepMathPhys:1991}.
Note that, despite their differences, the linear and von Neumann entropies are
going to show a similar trend as the quantum system evolves, which is also expected in the case
of their classical counterparts if there is a proper correspondence.
Furthermore, although the linear entropy is not an approximation of the von Neumann entropy,
it can readily be seen that, by Taylor expanding the latter, the first two terms correspond
to the former:
\be
 S_V =
- {\rm Tr} (\tilde{\rho} \ln \tilde{\rho}) =
- {\rm Tr} \left[ \tilde{\rho} \sum_{k=1}^\infty (-1)^{k+1}\ \frac{(\tilde{\rho} - \mathbb{I})^k}{k}
\right]
= 1 -  {\rm Tr} (\tilde{\rho}^2)
- {\rm Tr} \left[ \tilde{\rho} \sum_{k=2}^\infty (-1)^{k+1}\ \frac{(\tilde{\rho} - \mathbb{I})^k}{k}
\right] .
 \label{demo}
\ee
This thus shows that the von Neumann entropy partly contains basic information about the
system mixedness also provided by the linear entropy.

In the Wigner representation, the von Neumann entropy reads as
\begin{equation}
 S_V = - 2\pi\hbar \int \tilde{\rho}_W(x,p_x)
  \left( \ln \tilde{\rho} \right)_W (x,p_x) dx dp_x .
 \label{eq52}
\end{equation}
Note that, due to the non-commutativity of the logarithm operation and the Wigner transform,
$\left( \ln \tilde{\rho} \right)_W \ne \ln \tilde{\rho}_W$.
This avoids the drawback of the negativity displayed by the Wigner distribution function
$\tilde{\rho}_W$ in phase-space regions associated with interference.
In the classical case, as distributions are positive definite on any phase-space region,
it can be assumed that the equality holds.
Thus, we consider here the classical analog of (\ref{eq52}) as a Shannon-type entropy, henceforth
denoted as $S_V^{\rm cl}$, with functional form
\begin{equation}
 S_V^{\rm cl} =  - \int \tilde{\rho}_{\rm cl}(x,p_x) \ln \left[ 2\pi\hbar \tilde{\rho}_{\rm cl}(x,p_x) \right] dx dp_x .
 \label{eq51}
\end{equation}
Note in this expression the $2\pi\hbar$ factor inside the logarithm, necessary to obtain a
dimensionless argument.
In the quantum expression, Equation~(\ref{eq52}), it is outside the logarithm, because the density
matrix is dimensionless; in this case, the action dimensions are assigned to the full function
$\left( \ln \tilde{\rho} \right)_W$.
Consequently, although the quantum and classical expressions for the linear entropy keep a close
correspondence, the same does not hold for the von Neumann entropy, which is directly connected to
the third term on the r.h.s.\ of the second equality in Equation~(\ref{demo}).
Hence, for instance, in the case of quantum and classical states described in phase space by a
Gaussian distribution, the linear entropy is zero in both cases, while the von Neumann entropy
only vanishes for the quantum state; the ``classical'' von Neumann entropy renders a value of
$0.307$ (see Section~\ref{sec3-2}).


\section{Results and Discussion}
\label{sec3}


\subsection{Working System and Dynamical Evolution}
\label{sec3-1}

\begin{figure}[t]
 \centering
 \includegraphics[width=0.66\textwidth]{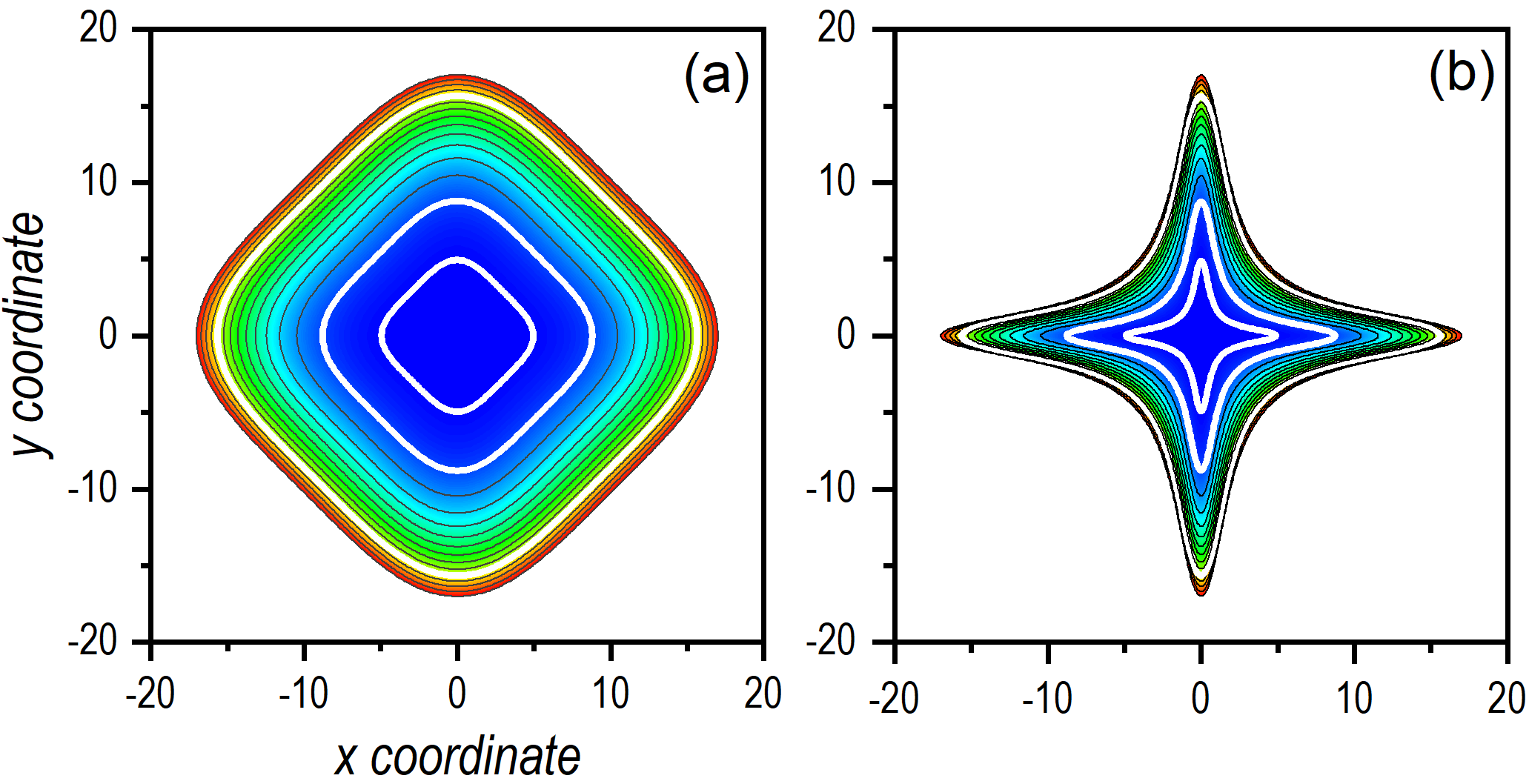}
 \caption{\label{fig1}
  {Density} plots of the quartic potential for $\beta = 0.01$ and two values of $\alpha$ that
  generate different dynamics: (\textbf{a}) $\alpha = 0.03$ (regular dynamics) and (\textbf{b}) $\alpha = 1$
  (chaotic dynamics).
  From blue to red, increasing value of the potential function, black contours indicate
  equipotential lines at multiples of $E=15$ (last contour shown corresponds to $E=210$).
  White contours in both panels represent the equipotential lines for $E_0=1.5, 15$, and $150$,
  which are the energies considered in the numerical simulations here (see next sections).}
\end{figure}

The working system here consists of two identical one-dimensional quartic oscillators coupled by
a nonlinear term, described by the Hamiltonian
\begin{equation}
 H(x,y,p_x,p_y) = \frac{p_x^2}{2m} + \frac{p_y^2}{2m} + \frac{\beta}{4} \left( x^4 + y^4 \right)
 + \frac{\alpha}{2}\ x^2 y^2 .
 \label{eq57}
\end{equation}
Due to its connection to chaotic dynamics, this model has been extensively studied in the
literature both classically \cite{HDmeyer:JCP:1986,dahlqvist:PRL:1990,bohigas:PhysRep:1993}
and also quantum mechanically
\cite{eckhardt-pollak:PRA:1989,bohigas:PhysRep:1993,joy:ModPhysLettB:1993,polavieja-borondo:PRL:1994}.
In particular, the following parameter values have been considered: $m = 1$, $\beta = 0.01$, and, for
the quantum case, $\hbar = 1$, all in arbitrary units.
Regarding the $\alpha$ parameter, which controls the dynamics, two values have been chosen, namely,
$\alpha = 0.03$, which produces regular dynamics, and $\alpha = 1$, which leads to fully chaotic
dynamics.
The two cases are illustrated in the density plots displayed in Figure~\ref{fig1}, where black
solid lines denote contours separated by multiples of 15 energy units, from 0 to a maximum energy
value of 210.
The white solid contours denote the three energies considered in the calculations described in
next sections $E_0 = 1.5, 15$, and $150$.
As it is clearly seen, changing from $\alpha = 0.03$ to $\alpha = 1$ leads to a dramatic change
of the equipotentials, which translates into passing from regular dynamical regimes to chaotic
ones.

The numerical simulations have been carried out as follows.
Quantum mechanically, the split-operator algorithm has been used to solve the time-dependent
Schr\"odinger equation \cite{sanz-bk-2}.
At each time-step, the density matrix has been computed, its trace over the subsystem $Y$ has been
performed (integrating over the $y$-coordinate), and then the reduced density matrix for the subsystem
$X$ has been numerically diagonalized, from which both the linear and von Neumann entropies have
been obtained.
Note that, from a practical point of view, there is no need to formerly determine the Wigner
function, because the interest relies directly on the actual values of the entropy and not on the
explicit form of the corresponding Wigner distribution function, thus being independent of the
representation considered.
This procedure thus avoids an unnecessarily extra increase in the computational demand involved
in the calculation of the four-dimensional Wigner distribution function for the full system.
As for the choice of the initial states, they are described in more detail in the next sections,
in the context of the actual computations.
In all cases considered, i.e., Gaussian states, cat states, and Bell-type entangled states, the
width of the corresponding wave packets involved has been chosen to be $\sigma^2 = 0.5$.

Concerning the classical calculations, they are based on performing statistics over trajectories
that obey the Hamilton equations of motion corresponding to the Hamiltonian (\ref{eq57}), with
their initial conditions selected according to the corresponding classical distribution (see
next sections), thus corresponding to an average energy around a certain value $E_0$.
As the potential function is homogeneous, the classical dynamics is energy-scale invariant,
i.e., for $\alpha$ and $\beta$ fixed, the same dynamics is observed at any value of $E_0$.
However, the statistics change, because the size and shape of the initial distribution of the
swarm of trajectories also play a role.
Moreover, the scale-invariance does not apply either to the quantum dynamics, where the larger
$E_0$, the larger the number of eigenstates involved in the dynamics
\cite{polavieja-borondo:PRL:1994}.
The statistics have been carried out over a total of $10^6$ trajectories by means of a box-counting procedure, which renders a fair coarse-grained value of the corresponding entropies
\cite{brumer-hamilton:PRA:1982,brumer-hamilton:JCP:1983,brumer-christoffel:PRA:1986}.
In fact, to establish a better comparison between the classical and quantum results, the pixel
size of the reduced phase space (i.e., the phase space for the $X$ subsystem, determined by $x$
and $p_x$), such box-counting has been performed taking into account the grid size associated
with the quantum partner.
Once the classical distribution is reconstructed at each time, in terms of a discretized phase space,
the classical linear and von Neumann entropies are straightforwardly computed by also discretizing
the corresponding trace operations (i.e., summing up over pixels).


\subsection{Gaussian State}
\label{sec3-2}

Let us first consider that both subsystems are initially described by localized Gaussian states
in the phase space, around the points $(x_0,p_{x,0})$ and $(y_0,p_{y,0})$, such that they are bound
by the energy conservation condition
\begin{equation}
 E_0 = \frac{p_{x,0}^2}{2m} + \frac{p_{y,0}^2}{2m} + \frac{\beta}{4} \left( x_0^4 + y_0^4 \right)
 + \frac{\alpha}{2}\ x_0^2 y_0^2 .
 \label{eq57E0}
\end{equation}
The joint state is given by the separable wave function
\be
 \Psi(x,y) = \left( \frac{1}{2\pi\sigma^2} \right)^{1/2}
  e^{-(x-x_0)^2/4\sigma^2 + ip_{x,0} \!\ x/\hbar}
  e^{-(y-y_0)^2/4\sigma^2 + ip_{y,0} \!\ y/\hbar} .
 \label{eq24}
\ee
Due to its factorizability, the corresponding Wigner distribution function is simply the product
of the reduced Wigner distribution functions associated with each subsystem:
\be
 \rho_W(x,y,p_x,p_y) = \left( \frac{1}{\pi\hbar} \right)^2
  e^{-(x - x_0)^2/2\sigma^2 - 2\sigma^2(p_x - p_{x,0})^2/\hbar^2}
  e^{-(y - y_0)^2/2\sigma^2 - 2\sigma^2(p_y - p_{y,0})^2/\hbar^2} .
 \label{eq25}
\ee
Thus, tracing over $Y$ has no influence on the reduced Wigner distribution function for $X$,
which reads as
\begin{equation}
 \tilde{\rho}_W(x,p_x) = \left( \frac{1}{\pi\hbar} \right)
  e^{-(x - x_0)^2/2\sigma^2 - 2\sigma^2(p_x - p_{x,0})^2/\hbar^2} .
 \label{eq26}
\end{equation}
From this expression, it is readily shown that the purity satisfies the above-mentioned unitarity
property, i.e.,
\begin{equation}
 {\rm Tr}_X [ \tilde{\rho}_W^2 ] =
  2\pi\hbar \int \tilde{\rho}_W^2 (x,p_x) dx dp_x = 1 ,
 \label{eq27}
\end{equation}
and hence the linear entropy, $S_L$, vanishes.
This result is valid for both the quantum and the classical case.
The same, though, does not hold for the von Neumann entropy, which vanishes in the quantum
case, but its classical counterpart, Equation~(\ref{eq51}), renders
\be
 S_V^{\rm cl} = 1 - \ln 2 \approx 0.307 .
 \label{eq28}
\ee
This value constitutes a lower bound for the classical von Neumann entropy; at any other stage of its
evolution, governed by the Hamiltonian (\ref{eq57}), the system entropy will be larger regardless of
the dynamics exhibited (either regular or chaotic).
This lower bound for the classical von Neumann entropy can be associated with the lack of coherence of
classical distributions, which not only prevents them from developing and displaying interference traits,
but also to compress and minimize the amount of phase-space information contained in them.
In other words, while a quantum pure state provides us with full information about the system (we cannot
further specify the state beyond the information encoded in its density matrix), classically a statistical
sampling of incoherent phase-space points (swarm of independent trajectories) is required to determine
the system state.

\begin{figure}[t]
 \centering
 \includegraphics[width=0.99\textwidth]{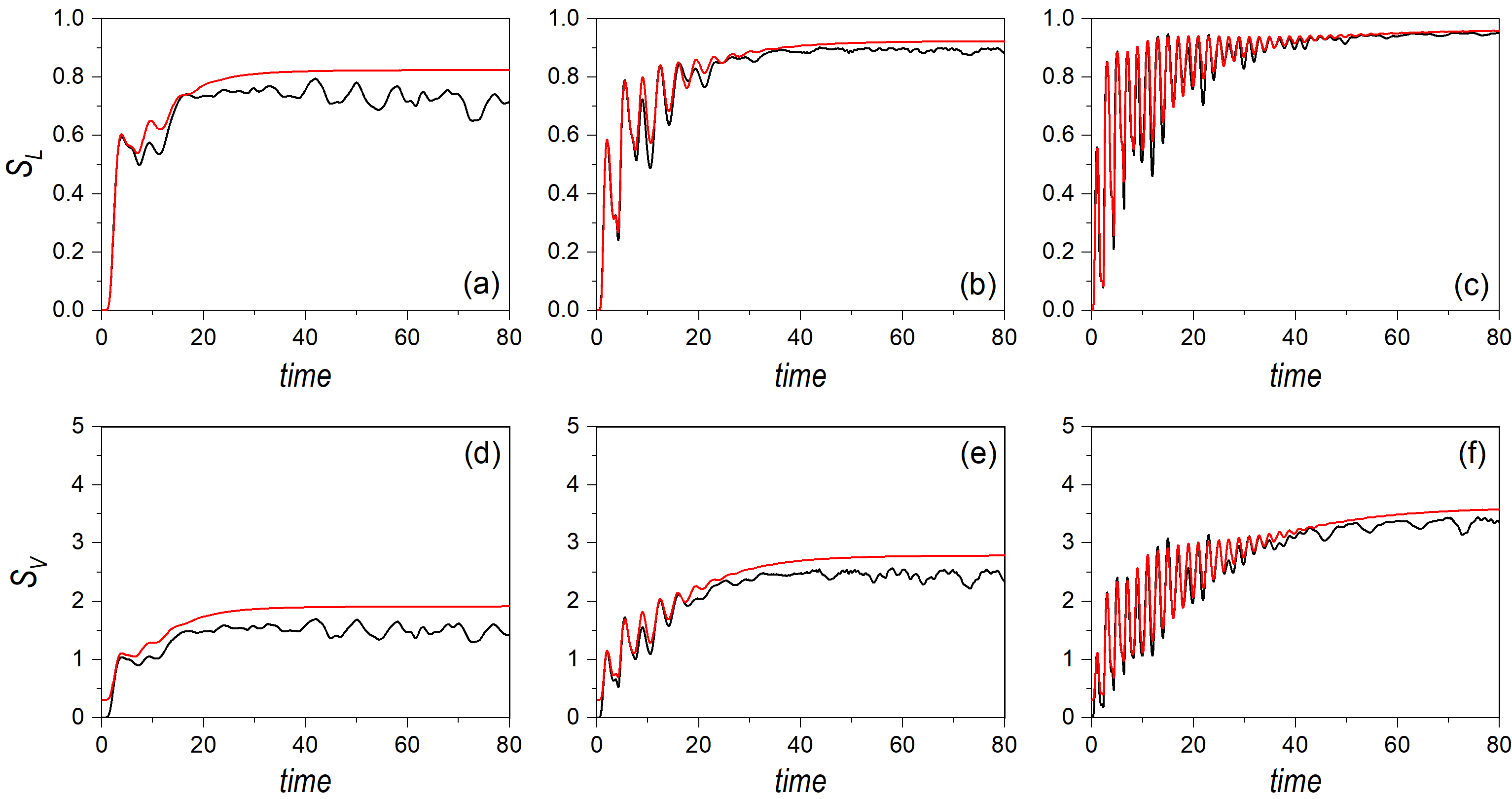}
 \caption{\label{fig2}
  Linear entropy (upper row panels) and von Neumann entropy (lower row panels) for an
  initial distribution associated with a single Gaussian wave packet at three different
  energies: $E_0 = 1.5$ (\textbf{a},\textbf{d}), $E_0 = 15$ (\textbf{b},\textbf{e}), and $E_0 = 150$ (\textbf{c},\textbf{f}).
  The dynamics corresponds to regularity conditions, with $\alpha = 0.03$.
  In all panels, the black line denotes the quantum results, while the red line represents the
  classical ones.}
\end{figure}

\begin{figure}[t]
	\centering
	\includegraphics[width=0.99\textwidth]{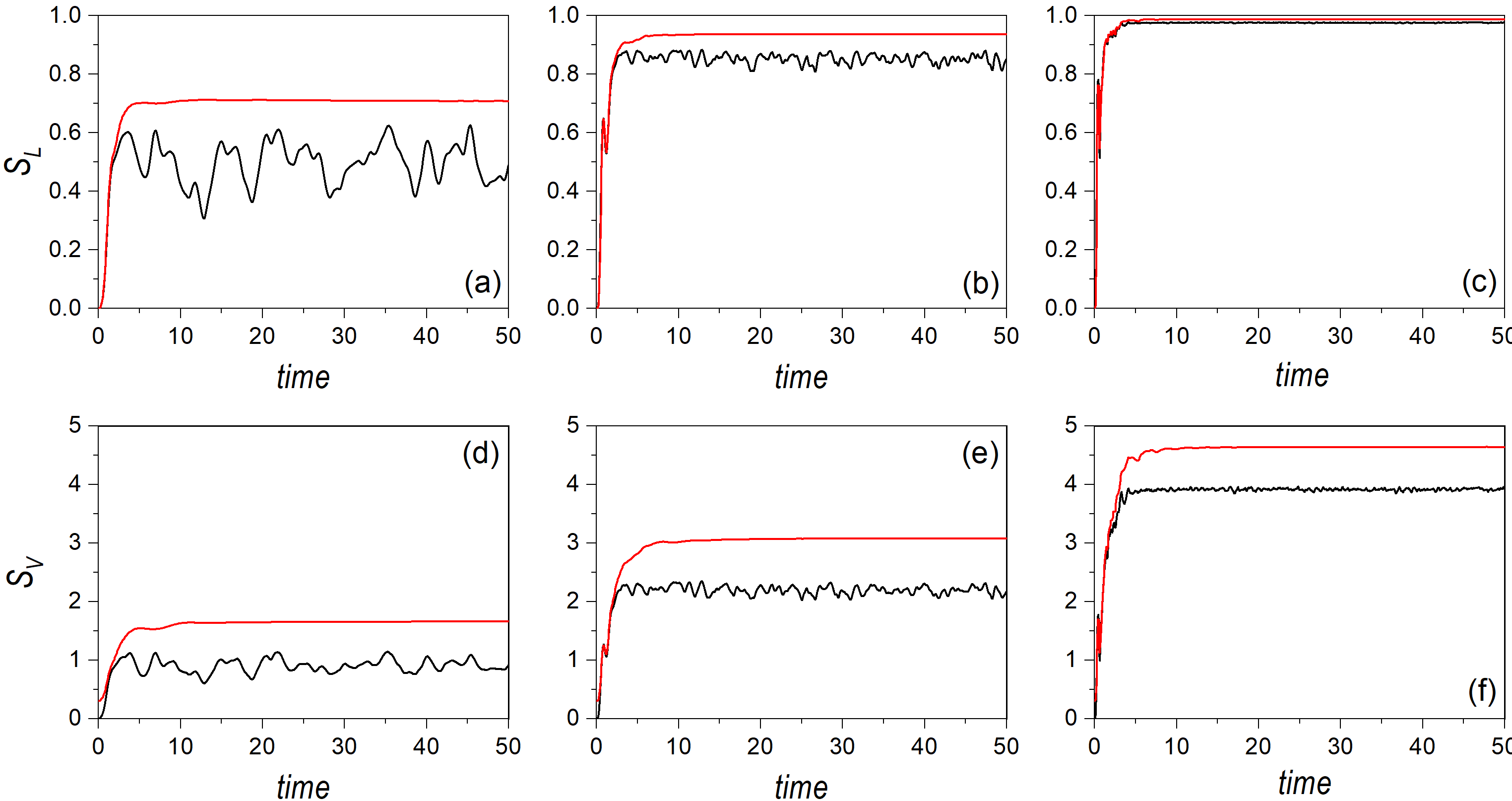}
	\caption{\label{fig3}
		Linear entropy (upper row panels) and von Neumann entropy (lower row panels) for an
		initial distribution associated with a single Gaussian wave packet at three different
		energies: $E_0 = 1.5$ (\textbf{a},\textbf{d}), $E_0 = 15$ (\textbf{b},\textbf{e}), and $E_0 = 150$ (\textbf{c},\textbf{f}).
		The dynamics corresponds to chaos conditions, with $\alpha = 1$.
		In all panels, the black line denotes the quantum results, while the red line represents the
		classical~ones.}
\end{figure}

If the Gaussian distributions are left to evolve according to the dynamics ruled by
the Hamiltonian (\ref{eq57}), we find that the above analytical results are in agreement
with the numerical simulations at the early stages of the evolution, as can be seen
in all cases represented in Figures~\ref{fig2} and~\ref{fig3}.
In these figures, the linear entropy is shown in the upper row panels, while the von Neumann
entropy is displayed in the lower row ones for $\alpha = 0.03$ (Figure~\ref{fig2}) and
$\alpha = 1$ (Figure~\ref{fig3}).
In all cases, the phase-space point around which the Gaussian distribution is centered corresponds to
$(0,0,\sqrt{mE_0},\sqrt{mE_0})$, i.e., the periodic orbit moving along the diagonal $y = x$
\cite{eckhardt-pollak:PRA:1989}, with $E_0 = 1.5$ (a/d), $E_0 = 15$ (b/e), and $E_0 = 150$ (c/f).
In Figure~\ref{fig2}, it can be noticed in that, under regular conditions, the classical entropies follow
very closely the behavior displayed by their quantum counterparts, with a better agreement, indeed, at
larger energies, where the amount of eigenstates involved in the quantum dynamics is larger.
Actually, the oscillations (recurrences) undergone by both entropies are nicely followed by the
classical distribution, even though one might expect important interference-mediated differences.
This trend is more clearly seen (for a longer time) as the frequency of the periodic orbit becomes larger,
which happens as $E_0$ increases.
A rather simple estimation suffices to show that, classically, the half-period, $\tau_{1/2}$, associated
with these orbits is a function of $E_0$:
\be
 \tau_{1/2} (E_0) = \frac{1}{\sqrt{mE_0}} \int_{x_-}^{x_+} \frac{dx}{\sqrt{ 1 - \gamma x^4}}
  = \frac{2 x_+}{\sqrt{mE_0}}\ {_2}F_1(1/4,1/2;5/4; x_+^4 \gamma)
  \approx 3.12 \left[ m^2 (\beta + \alpha) E_0 \right]^{-1/4} ,
\ee
where $x_\pm = \pm \gamma^{-1/4}$ denote the turning points of the periodic orbit, with
$\gamma = (\beta + \alpha)/2E_0$, and ${_2}F_1(a,b;c;z)$ is the Gaussian or ordinary hypergeometric
function of the first kind, with ${_2}F_1(1/4,1/2;5/4;1) \approx 1.31$.
The values of the half period corresponding to the three energies here considered are:
$\tau_{1/2}(1.5) \approx 6.30$, $\tau_{1/2}(15) \approx 7.1$, and $\tau_{1/2}(150) \approx 1.99$.
As the Gaussian distribution is launched from the center of the potential, it will undergo
a recurrence at half the period, each time that the classical periodic orbit crosses this point
(maxima indicate maximum amplitude in the coordinate, and minima, maximum amplitude in the
momentum).

In Figure~\ref{fig2}a, however, recurrences are hardly seen, which is due to the smearing out undergone
by the distribution, as the back-and-forth oscillatory motion of the Gaussian is slower than its
spreading.
To better understand this effect, consider a free Gaussian wave packet.
Its spreading in time is determined by the simple expression \cite{sanz:JPA:2008,sanz-bk-2}
\be
 \sigma(t) = \sigma \sqrt{ 1 + \left( \frac{\hbar t}{2m\sigma^2} \right)^2 } .
\ee
In the particular case of the bound potential here considered, we can assume that there is an upper bound
for the time at which the width of the wave packet will cover the extension between the two classical
turning points,
\be
 \Delta u = x_+ - x_- = 2 \left( \frac{2mE_0}{\beta + \alpha} \right)^{1/4} .
\ee
Accordingly, if $\Delta u \approx \sigma(t)$, we obtain that $\Delta u(1.5) \approx 4$.
That is, by the time that it takes the system to pass through its initial position for the first time, the
corresponding distribution has already covered nearly the whole available space, thus making difficult to
detect subsequent recurrences, as it can be noticed in Figure~\ref{fig2}a,d.
Nonetheless, due to interference, the phase space cannot be homogeneously covered, as it happens with
the swarms of trajectories, due to ergodicity. 
Hence, it can be seen that the quantum entropies display a series of fluctuations that are absent in the
classical partners, which approach the saturation limit in a smooth manner.
On the contrary, at higher energies, the spreading times are shorter than the orbit half periods, and
hence more recurrences can be observed, which has nothing to do with interference, as the classical
system also exhibits such recurrences.
Thus, this can be regarded as a purely classical effect, which is enhanced as $E_0$ becomes larger.
As it is seen in Figure~\ref{fig2}c,f, for $E_0=150$, these tiny oscillations associated with the
classical periodic orbit extend for a longer time.
In this case, only after $t \approx 40$, the effects due to interference, associated with the second
(coherence) term of the Moyal bracket, start being noticeable, particularly in the von Neumann entropy---note that these effects are more apparent in the von Neumann entropy than in the linear entropy due
to the third term in Equation~(\ref{demo}).
Roughly speaking, it can be seen that the lifetime 
of the oscillatory part of both entropies doubles each
time the energy increases in one order of magnitude, which is also the trend found with the spreading time:
$t(1.5) \approx 4$, $t(15) \approx 7.3$, and $t(150) \approx 13.1$.

In the case of an underlying classically chaotic dynamics, when $\alpha = 1$ (see Figure~\ref{fig3}), both
the periods of the classical orbits and the spreading times become shorter, which basically turns into a
fast suppression of the initial oscillatory behavior of the entropies.
Interference terms are thus more prominent in the quantum entropies, particularly at low energies,
when the number of eigenstates involved is relatively low, while the classical counterparts
quickly reach the saturation regime (the trajectories tend to cover the whole phase space).
As the energy increases, though, the fluctuations associated with interference become faster
and with a smaller amplitude, which makes the quantum entropies to display a sort of saturation
regime similar to the classical one.
Indeed, it can be noticed that the difference between the quantum and classical linear entropies
becomes negligible, with their saturation value nearly reaching the maximum.
This indicates a high effective production of entanglement between $X$ and $Y$, which eventually leads to
maximal mixing in both subsystems' states.
Regarding the von Neumann entropies, the trend is analogous to the linear counterparts, with
the relative distance between the quantum and classical average saturation values,
$|S_V^{\rm cl} - S_V|/S_V^{\rm cl}$, becoming smaller.

It is also worth emphasizing that, due to the range of energies considered, it is possible
to detect a dynamical crossover concerning the generation of entanglement in the reference system.
Comparing Figures~\ref{fig2} and \ref{fig3}, particularly the plots of the von Neumann entropies
(although the linear entropies also show the same trend), it is readily noticed that at low
energies the production of entanglement is more effective in the regular regime than in the
chaotic one, both classically and quantum mechanically.
Observe the asymptotic trends, that is, the maximum for the classical dynamics, and the average
for the quantum mechanical one.
These values are lower in the chaotic regime than in the regular one.
At intermediate energies, both dynamics are similar, although the distance between the quantum
and classical asymptotic trends is larger in the chaotic case than in the regular one.
However, at high energies, we observe a crossover; the production of entanglement becomes more
effective in the chaotic case than in the regular one.
This is a counter-intuitive effect, against a common statement that underlying classically chaotic dynamics
should generate entanglement more efficiently, but that confirms analogous results found in many-body
interactions at different thermal regimes (and hence different dynamical behaviors) \cite{sanz:PRE:2012}.

\begin{figure}[t]
 \centering
 \includegraphics[width=0.66\textwidth]{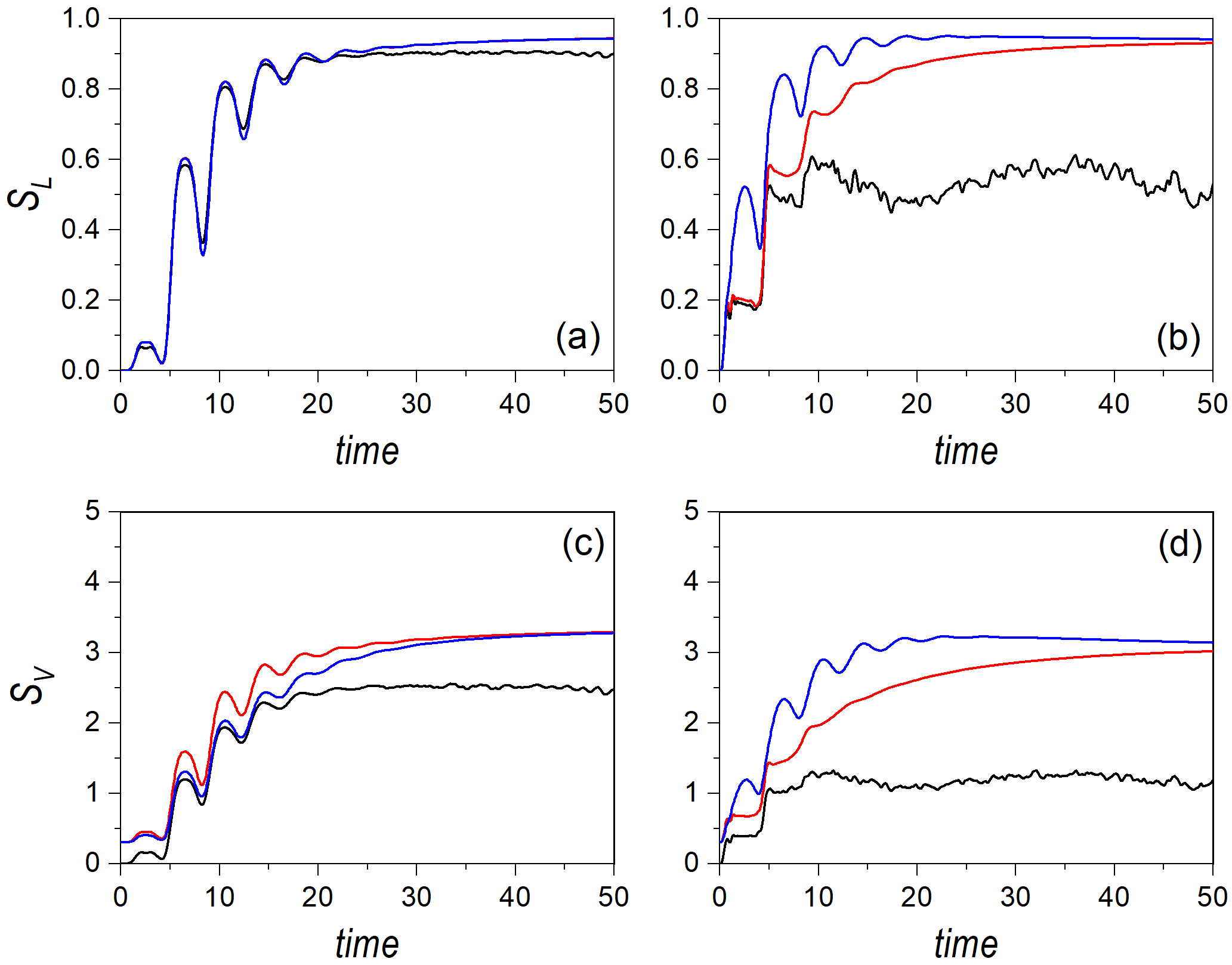}
 \caption{\label{fig4}
 Linear entropy (upper row panels) and von Neumann entropy (lower row panels) for an initial
 distribution associated with a single Gaussian wave packet moving along the channel
 at an energy $E_0 = 15$, and affected by regularity (\textbf{a},\textbf{c}) and chaos (\textbf{b},\textbf{d}) conditions.
 In all panels, the black line denotes the quantum results, the blue line represents classical
 motion initially oriented along the $x$-direction, and the red line classical motion initially
 directed along the $y$-direction.}
\end{figure}

In order to further explore these aspects, let us now consider the case where the Gaussian
distribution is launched, again, from the center of the potential, but directed along one of the
transverse directions, either the $x$-direction, initially centering the Gaussian on
$(0,0,\sqrt{2mE_0},0)$, or the $y$-direction, considering it on the alternative phase-space point
$(0,0,0,\sqrt{2mE_0})$.
The corresponding linear and von Neumann entropies are respectively displayed in the upper and lower
row panels of Figure~\ref{fig4} for $E_0=15$.
Furthermore, in order to better evaluate the effects of the coupling strength, $\alpha$, the results
for the regular and the chaotic cases are shown, respectively, in panels (a/d) and (b/c).
As it can be noticed, the quantum entropies do not discriminate the initial direction of the motion,
due to the symmetry of the two selected directions.
That is, it does not matter who is initially at rest, either $Y$ or $X$, assigning all the initial
energy, $E_0$, to the other subsystem.
As mentioned above, in Section~\ref{sec2}, because the joint state describing both systems is pure, both entropy measures render the same result for the two subsystems, and hence a symmetric exchange of
the momentum does not produce any change in the respective entropies.
The same, however, is not true in the classical case, where we can observe major differences between
both directions until the entropies reach the saturation regime, regardless of the dynamical regime
considered.
This is consistent with the fact that the trajectories initially started with some momentum along the
$x$-direction must, somehow, manifest the oscillatory behavior along this direction, with small
deviations towards the $y$-direction, as $p_{y,0}=0$.
In fact, following the above procedure to determine the half-period, we find that
\be
 \tau'_{1/2} (E_0) = \frac{1}{\sqrt{2mE_0}} \int_{x'_-}^{x'_+} \frac{dx}{\sqrt{ 1 - \gamma' x^4}}
 \approx 2.62 \left( m^2 \beta E_0 \right)^{-1/4} ,
\ee
where $x'_\pm = \pm \gamma'^{-1/4}$ denote the actual turning points of the periodic orbit along the
$x$-direction, with $\gamma' = \beta/4E_0$.
As, now, there is no dependence on the coupling strength $\alpha$, the same half-period,
$\tau'_{1/2} (15) \approx 4.21$, is obtained both in the regular and in the chaotic dynamics.
This is clearly seen in all panels of Figure~\ref{fig4} by inspecting the blue line during the first
stages of the volution (up to $t \approx 20$).

There are, though, important differences between the regular and chaotic dynamics, in particular, the
observation of oscillatory motion also along the $y$-direction in the former and its absence (but with
presence of kind of steps of the same duration) in the latter.
This lack in the chaotic regime arises because the interaction potential acquires the shape of
two crossed channels (see Figure~\ref{fig1}b), which prevents the motion along the corresponding
perpendicular direction (here, for $p_{x,0}=0$, along the $x$-direction) before the swarm of trajectories
has smeared out across the available phase space.
On the contrary, in the regular regime, the seemingly squared-billiard shape displayed by the interaction
potential (see Figure~\ref{fig1}a) allows an effective energy transfer from one degree of freedom to the
other, as in an oscillator.
When these two behaviors are compared with the quantum result, we find that the latter resembles the case
of motion along the $x$-direction in the regular regime, while it draws closer to the case of motion along
the $y$-direction in the chaotic regime.
This has to do with the capability of the Gaussian distribution to spread across the potential well,
which is more efficient in the regular case than in the chaotic one, due to the same reason
mentioned above regarding the diffusion of classical trajectories.

\begin{figure}[t]
 \centering
 \includegraphics[width=0.66\textwidth]{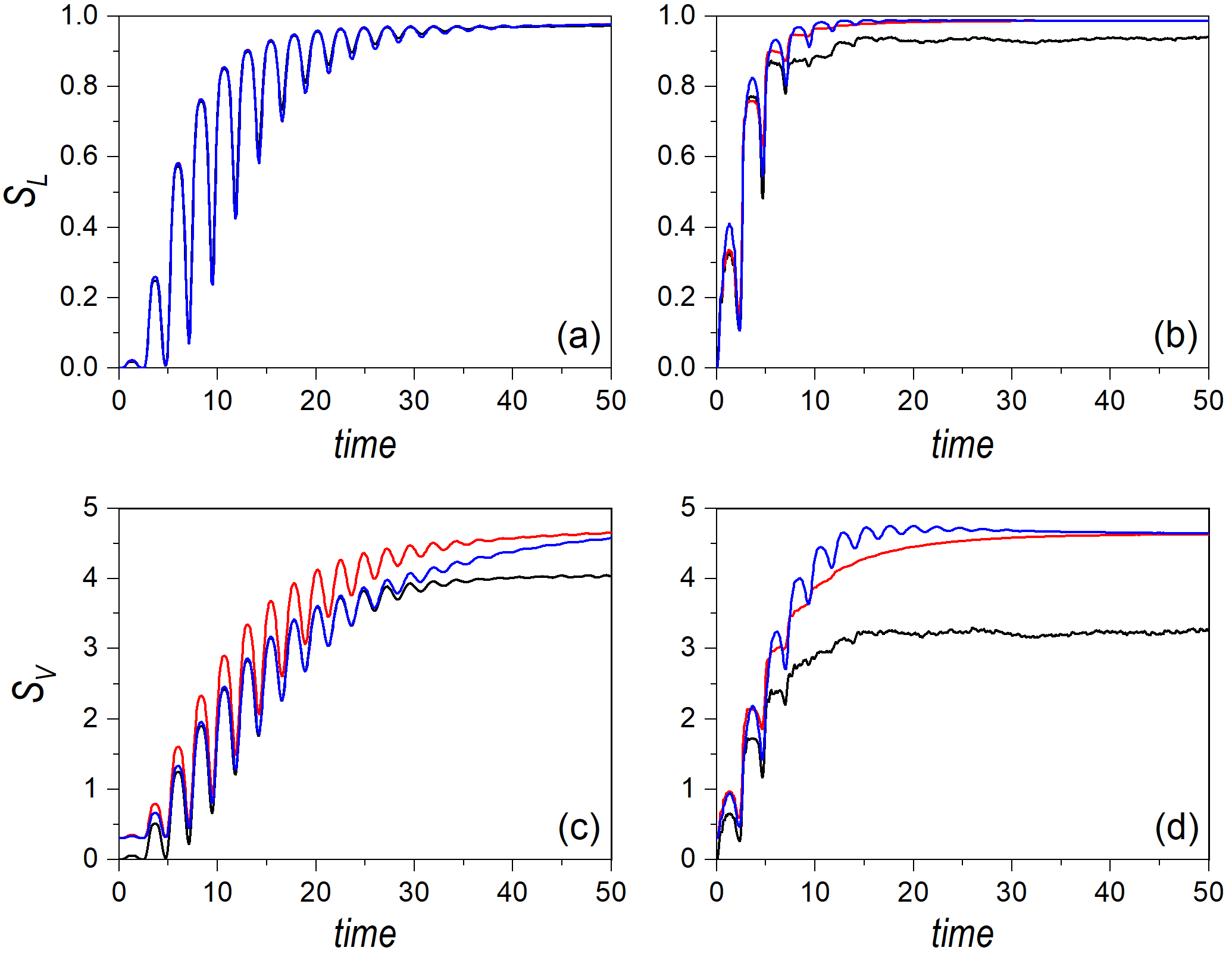}
 \caption{\label{fig5}
 Linear entropy (upper row panels) and von Neumann entropy (lower row panels) for an initial
 distribution associated with a single Gaussian wave packet moving along the channel
 at an energy $E_0 = 150$, and affected by regularity (\textbf{a},\textbf{c}) and chaos (\textbf{b},\textbf{d}) conditions.
 In all panels, the black line denotes the quantum results, the blue line represents classical
 motion initially oriented along the $x$-direction, and the red line classical motion initially
 directed along the $y$-direction.}
\end{figure}

That situation changes if higher energies are considered, as is shown in Figure~\ref{fig5},
for $E_0=150$, and where each panel represents the same quantities and conditions as in Figure~\ref{fig4}.
In this case, the half-period reduces to $\tau'_{1/2} (150) \approx 2.37$, which allows us to observe
a larger number of recurrences in the entropies.
This effect is even more remarkable in the chaotic case, as, as seen above for $E_0=15$, such
oscillatory behavior is absent in the same regime.
Furthermore, unlike the distribution moving along the diagonal, here the production of entanglement is
more efficient in the regular case, although the difference between the corresponding entropies becomes
smaller as $E_0$ increases, which is consistent with the previous results.
Comparing Figures~\ref{fig4} and~\ref{fig5}, a clear crossover trend is also apparent, as before,
although it will take place at higher energies.
This can be associated with the fact that, in this case, in the chaotic regime, the motion is confined
within a channel for a longer time (that is, it takes longer to spread out within the well), thus
displaying a lower rate of ergodicity.
This is confirmed by the fact that the (asymptotic) maximum value of the entropy is lower than in the
case of diagonal motion analyzed above (compare Figures~\ref{fig4} and~\ref{fig5} with the analogous cases in Figures~\ref{fig1} and~\ref{fig2}).
Thus, we can conclude that the crossover is a general trend, although, like the Lyapunov exponent
in a classical dynamical system, it depends on the specific phase-space point where the quantum
distribution is launched from, as well as the energy $E_0$ at which this is done.

As we will see next, this dynamical analysis already renders some interesting insights to understand
the behaviors displayed by cat states and Bell-type entangled states, as well as the suitability of the
classical counterparts here defined.

\subsection{Cat State}
\label{sec3-3}

Gaussian states are known for having a direct classical counterpart, although we have also seen above
that there are important dynamical differences.
A natural question that now arises is whether similar trends and agreements can also be found beyond
these particular states.
Thus, let us consider the non-classical case of an initial coherent superposition for $X$, consisting
of two Gaussian wave packets centered at the phase-space points $(x_1, p_{x,1})$ and $(x_2, p_{x,2})$,
coupled to $Y$, which is still described by a single Gaussian centered at $(y_0, p_{y,0})$
(as described in Section~\ref{sec3-2}).
The joint state reads as
\be
 \Psi(x,y) = \frac{A}{\sqrt{2}} \left(\frac{1}{2\pi\sigma^2}\right)^{1/2}
  \left[ e^{-(x-x_1)^2/4\sigma^2 + ip_{x,1}\!\ x/\hbar}
   + e^{-(x-x_2)^2/4\sigma^2 + ip_{x,2}\!\ x/\hbar} \right]
  e^{-(y-y_0)^2/4\sigma^2 + ip_{y,0}\!\ y/\hbar} ,
 \label{eq30}
\ee
with
\be
 A^{-1} = \sqrt{ 1 + e^{-(\Delta x)^2/8\sigma^2 - \sigma^2 (\Delta p)^2/2\hbar^2}
  \cos \left( \Delta p_x \bar{x}/\hbar \right) } ,
 \label{eq31}
\ee
which amounts to 1 if the distance in phase space between the two wave packets is large enough, and
where $\bar{x} = (x_1 + x_2)/2$, $\Delta x = x_1 - x_2$, and $\Delta p_x = p_{x,1} - p_{x,2}$.
The corresponding Wigner distribution function is given by
\begin{eqnarray}
 \rho_W(x,y,p_x,p_y)\!\!\!\! & = & \!\!\!\! \frac{A^2}{2} \left(\frac{1}{\pi\hbar}\right)^2 \
  \bigg\{ e^{-(x-x_1)^2/2\sigma^2 - 2\sigma^2(p_x - p_{x,1})^2/\hbar^2}
  + e^{-(x-x_2)^2/2\sigma^2 - 2\sigma^2(p_x - p_{x,2})^2/\hbar^2}
  \nonumber \\
  \!\!\!\! & & \!\!\!\! \quad + 2 e^{-(x - \bar{x})^2/2\sigma^2 - 2\sigma^2(p_x - \bar{p}_x)^2/\hbar^2}
  \cos \left[\frac{\Delta p_x x - (p_x - \bar{p}_x) \Delta x}{\hbar}\right] \bigg\} ,
  \nonumber \\ \!\!\!\! & & \!\!\!\!
  \qquad \times e^{-(y-y_0)^2/2\sigma^2 - 2\sigma^2(p_y - p_{y,0})^2/\hbar^2} ,
 \label{eq32}
\end{eqnarray}
with $\bar{p}_x = (p_{x,1} + p_{x,2})/2$.
Again, due to the separability of this joint state, integrating over the $Y$-system variables has no
influence on the reduced Wigner distribution function for $X$, which reads as
\begin{eqnarray}
 \tilde{\rho}_W(x,p_x) \!\!\!\! & = & \!\!\!\! \frac{A^2}{2\pi\hbar} \
  \bigg\{ e^{-(x - x_1)^2/2\sigma^2 - 2\sigma^2(p_x - p_{x,1})^2/\hbar^2}
  + e^{-(x - x_2)^2/2\sigma^2 - 2\sigma^2(p_x - p_{x,2})^2/\hbar^2}
  \nonumber \\
  \!\!\!\! & & \!\!\!\! + 2 e^{-(x - \bar{x})^2/2\sigma^2 - 2\sigma^2(p_x - \bar{p}_x)^2/\hbar^2}
  \cos \left[\frac{ \Delta p_x x - (p_x - \bar{p}_x) \Delta x }{\hbar}\right] \bigg\} .
 \label{eq33}
\end{eqnarray}
%

Unlike (\ref{eq25}), note that now the Wigner distribution function (\ref{eq32}) is negative definite
on certain phase-space regions (actually in the subspace corresponding to the subsystem $X$, as seen in
(\ref{eq33})), which is a distinctive trait of the mutual coherence between the two wave packets.
This negativity makes the correspondence with a classical counterpart unclear.
A way to proceed could be assigning a negative flux to classical initial conditions
(trajectories) picked up within the negative regions.
However, the physical meaning of classical density distribution would get lost, and the
Shannon entropy could not be computed in the way described above.
Instead, another procedure has been chosen.
In (\ref{eq33}), there are three terms.
Two of them correspond to having the system localized either around the phase-space points ($x_1,p_1$) and
($x_2,p_2$), according to identical Gaussian distributions.
The third term, the interference one associated with the negativity of the Wigner distribution function,
is a cosine modulated by a Gaussian identical to the two previous ones, but centered around the mean values
$\bar{x}$ and $\bar{p}_x$.
As this is the contribution that makes the Wigner distribution function to be non-classical, let us
simply remove it and analyze its dynamical behavior in terms of the linear and von Neumann entropy
measures.
Let us thus assume that the ``classical'' cat state is a bare sum of localized Gaussian distributions,
such that
\ba
 \rho_{\rm cl}(x,y,p_x,p_y) \!\!\!\! & = & \!\!\!\!
 \frac{1}{2} \left(\frac{1}{\pi\hbar}\right)^2 \
 \bigg[ e^{-(x - x_1)^2/2\sigma^2 - 2\sigma^2(p_x - p_{x,1})^2/\hbar^2}
     + e^{-(x - x_2)^2/2\sigma^2 - 2\sigma^2(p_x - p_{x,2})^2/\hbar^2} \bigg] \nonumber \\
 \!\!\!\! & & \!\!\!\! \qquad \qquad \qquad
  \times e^{-(y - y_0)^2/2\sigma^2 - 2\sigma^2(p_y - p_{y,0})^2/\hbar^2} ,
 \label{eq34}
\ea
while the reduced density distribution is simply given by
\begin{equation}
 \tilde{\rho}_{\rm cl}(x,p_x) = \frac{1}{2\pi\hbar} \
  \bigg[ e^{-(x - x_1)^2/2\sigma^2 - 2\sigma^2(p_x - p_{x,1})^2/\hbar^2}
  + e^{-(x - x_2)^2/2\sigma^2 - 2\sigma^2(p_x - p_{x,2})^2/\hbar^2} \bigg] ,
 \label{eq35}
\end{equation}
which lacks the prefactor $A^2$ precisely because there is no interference term.

Unlike the single Gaussian distribution, now none of the classical entropies are going to equal the
quantum mechanical counterparts for these cat states.
This time the linear entropy reads as
\be
 S_L^{\rm cl} = \frac{1}{2} \left[ 1 - e^{-(\Delta x)^2/4\sigma^2 - \sigma^2(\Delta p_x)^2/\hbar^2} \right] .
\label{eq36b}
\ee
Thus, if the distance in phase space is relatively large, this quantity amounts to
$S_L^{\rm cl} \approx 0.5$.
The fact that this quantity does not vanish can be interpreted as a consequence of having
the classical distribution delocalized between two different regions in phase space.
This will also have an influence on the classical von Neumann entropy, which can be
written as
\begin{eqnarray}
	S_V^{\rm cl} \!\!\!\! & = & \!\!\!\!
	 - \int \frac{1}{2} \ (\tilde{\rho}_{\rm cl, 1} + \tilde{\rho}_{\rm cl, 2})
	\ln \left[ \frac{1}{2} \ (\tilde{\rho}_{\rm cl, 1} + \tilde{\rho}_{\rm cl, 2}) \right]
	\nonumber \\ \!\!\!\! & \simeq & \!\!\!\!
	- \frac{1}{2} \int \tilde{\rho}_{\rm cl, 1} \ln \tilde{\rho}_{\rm cl, 1}
	- \frac{1}{2} \int \tilde{\rho}_{\rm cl, 2} \ln \tilde{\rho}_{\rm cl, 2}
	+ \frac{1}{2} \int (\tilde{\rho}_{\rm cl, 1} + \tilde{\rho}_{\rm cl, 2}) \ln 2 ,
	\label{entrop}
\end{eqnarray}
if the reduced Wigner distribution function (\ref{eq35}) is recast as
\be
 \tilde{\rho}_{\rm cl} (x,p_x) = \frac{1}{2}
 \left[ \tilde{\rho}_{\rm cl, 1}(x,p_x) + \tilde{\rho}_{\rm cl, 2} (x,p_x) \right]
\ee
and the assumption of a relatively long distance in phase space between the two Gaussians
is again considered.
Now, given the symmetry of the problem, the integrals for both distributions render the same
result, and hence the above expression can be written, without loss of generality, as
\be
 S_V^{\rm cl} \approx - \int \tilde{\rho}_{\rm cl, 1} \ln \tilde{\rho}_{\rm cl, 1}
   + \ln 2 \int \tilde{\rho}_{\rm cl, 1} = 1 ,
\label{eq59}
\ee
where (\ref{eq28}) has been used in the last equality.
It is thus seen that the von Neumann entropy for the classical distribution does not simply
doubles the result for a single Gaussian, but, due to the delocalization, the lower bound
amounts to 1.

The above differences between the classical and quantum entropies, once the interference term has been
removed in the quantum distribution in order to define the classical one, already provide us with some
clues as to why quantum entropies always remain lower than classical ones.
This was an aspect already mentioned in the case of the Gaussian distributions, namely, that quantum
distributions optimize the compression of phase-space information by including such an interference term; however, this is not obvious, because there was no such term included. 
Only as time proceeds and the interference starts developing does this become clearer.
On the contrary, because the two subsystems become more entangled with time, the systems' interference
traits are gradually suppressed, which eventually leads the quantum and classical linear entropies to draw
closer, and the von Neumann ones to reduce the distance between corresponding quantum and classical
saturation (asymptotic) values.
On the other hand, the choice for the classical partner of the quantum cat here will also be clearer in
next section, when we will see that the same functional form also arises in the case of the entangled
state.
This ``classical'' partner for the entangled state emerges, precisely, after the coupling (entanglement)
with the environment washes out the interference term.
This thus makes clear why the same delocalized classical distribution serves to set a comparison with
both the cat state (quantum delocalization before undergoing decoherence) and the Bell-type entangled
state (quantum delocalization after undergoing decoherence).

\begin{figure}[t]
 \centering
 \includegraphics[width=0.66\textwidth]{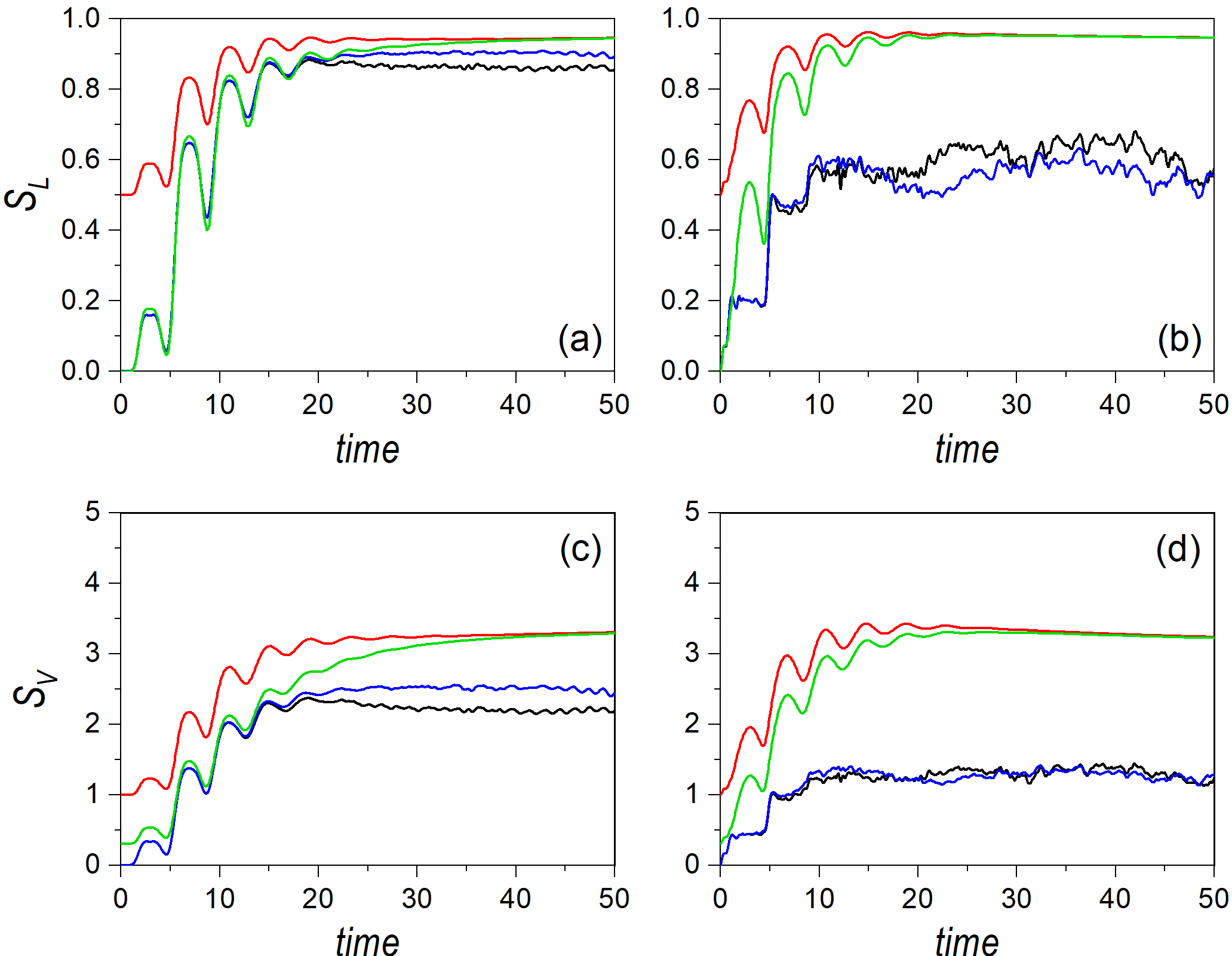}
 \caption{\label{fig6}
  Linear entropy (upper row panels) and von Neumann entropy (lower row panels) for an initial
  distribution associated with a cat state aligned along the $x$-channel,
  with an energy $E_0 = 15$, and affected by regularity conditions (\textbf{a},\textbf{c}) and chaos conditions (\textbf{b},\textbf{d}).
  In all panels, black line denotes the quantum results, while the red line refers to the classical ones.
  To compare with, the quantum and classical results for a single Gaussian distribution, with the initial
  at ${\bf p}_2$ (see text for details), are also included, and denoted, respectively, with the blue
  and green lines.}
\end{figure}

Such a trend is better understood by considering the dynamical evolution of the entropies.
Thus, consider a cat state formed by two Gaussian wave packets initially localized at the phase
space points ${\bf p}_1 = (-x_0,0,\sqrt{2mE'_0},0)$ and ${\bf p}_2 = (x_0,0,-\sqrt{2mE'_0},0)$, where
$x_0 = 2.5$ and $E'_0 = E_0 - \beta x_0^4/4$, with $E_0=15$.
This initial condition ensures a negligible overlapping.
The time evolution of the linear and von Neumann entropies are displayed, respectively, in the upper
and lower row panels of Figure~\ref{fig6}, in the left column for a regular dynamics ($\alpha = 0.03$)
and in the right column for the chaotic one ($\alpha = 1$).
Furthermore, to compare with, and hence to determine the relationship between the single, localized
Gaussian state and the delocalized cat state, the corresponding results for a single Gaussian centered
at ${\bf p}_2$ are also shown.
When comparing all these cases, several interesting features are worth stressing.
First of all, in all cases, except for some minor deviations, the quantum results for the cat state and
the single Gaussian distribution are identical; actually, in the chaotic regime, note that the results
mimic those of the Gaussian moving along the transverse channel (see Figure~\ref{fig4}).
This means that, quantum mechanically, the entropy measures do not make any distinction between having
a localized distribution in phase space or two separate distributions coherently connected by the
the interference term.
Classically, there is an initial difference, as was already discussed above, but, as time proceeds,
both distributions, localized and delocalized, approach the same asymtotic value due to ergodicity.
Furthermore, we can observe that the initial increase stage is characterized by the same
oscillatory behavior discussed in the previous section.
Notice that, although the entropies for the classical cat-state distribution start with non-vanishing
values, each Gaussian contributes independently due to their incoherence (the trajectories associated
with one of these Gaussians will display mirror symmetry with respect to the trajectories related to the
other Gaussian, thus rendering the same oscillation period and~behavior).

\subsection{Bell-Type Maximally Entangled State}
\label{sec3-4}

Finally, let us consider a Bell-type or maximally entangled state consisting of two delocalized
Gaussians for each subsystem, with centers $(x_1, p_{x,1})$ and $(x_2, p_{x,2})$ for $X$, and
$(y_1, p_{y,1})$ and $(y_2, p_{y,2})$ for $Y$, such that $(y_1, p_{y,1}) = (x_2, p_{x,2})$ and
$(y_2, p_{y,2}) = (x_1, p_{x,1})$:
\ba
 \Psi(x,y) \!\!\!\! & = & \!\!\!\! \frac{A}{\sqrt{2}} \left(\frac{1}{2\pi\sigma^2}\right)^{1/2} \
 \bigg[ e^{-(x - x_1)^2/4\sigma^2 + ip_{x,1} x/\hbar} e^{-(y - y_1)^2/4\sigma^2 + ip_{y,1} y/\hbar}
 \nonumber \\ \!\!\!\! & & \!\!\!\! \qquad
 + e^{-(x - x_2)^2/4\sigma^2 + ip_{x,2} x/\hbar} e^{-(y - y_2)^2/4\sigma^2 + ip_{y,2} y/\hbar} \bigg] ,
 \label{eq38}
\ea
where
\be
 A^{-1} = \sqrt{ 1 +
  e^{-(\Delta x)^2/8\sigma^2 - \sigma^2 (\Delta p_x)^2/2\hbar^2}
  e^{-(\Delta y)^2/8\sigma^2 - \sigma^2 (\Delta p_y)^2/2\hbar^2}
   \cos \left( \frac{ \Delta p_x \bar{x} + \Delta p_y \bar{y}}{\hbar} \right) } ,
 \label{eq39}
\ee
with $\bar{x}$, $\Delta x$, and $\Delta p_x$, and their homologs for $Y$, defined as before.
The associated Wigner distribution function is
\ba
 \rho_W(x,y,p_x,p_y) \!\!\!\! & = & \!\!\!\! \frac{A^2}{2} \left(\frac{1}{\pi\hbar}\right)^2
 \bigg\{ e^{-(x-x_1)^2/2\sigma^2 - 2\sigma^2(p_x - p_{x,1})^2/\hbar^2}
         e^{-(y-y_1)^2/2\sigma^2 - 2\sigma^2(p_y - p_{y,1})^2/\hbar^2}
 \nonumber \\ \!\!\!\! & & \!\!\!\! \qquad
      +  e^{-(x-x_2)^2/2\sigma^2 - 2\sigma^2(p_x - p_{x,2})^2/\hbar^2}
         e^{-(y-y_2)^2/2\sigma^2 - 2\sigma^2(p_y - p_{y,2})^2/\hbar^2}
  \nonumber \\ \!\!\!\! & & \!\!\!\! \qquad
  + 2 e^{-(x-\bar{x})^2/2\sigma^2 - 2\sigma^2(p_x - \bar{p})^2/\hbar^2}
      e^{-(y-\bar{y})^2/2\sigma^2 - 2\sigma^2(p_y - \bar{y})^2/\hbar^2}
  \nonumber \\ \!\!\!\! & & \!\!\!\! \qquad \quad
  \times \cos \left[\frac{ \Delta p_x\!\ x + \Delta p_y\!\ y
   - (p_x-\bar{p}_x) \Delta x - (p_y-\bar{p}_y) \Delta y }{\hbar}\right] \bigg\} ,
  \nonumber \\ \!\!\!\! & & \!\!\!\!
 \label{eq40}
\ea
which also contains an interference term that generates negative definite
phase-space regions, although now such regions are not embedded within a
particular subspace, but affect the full four-dimensional phase space.
Accordingly, because (\ref{eq40}) is not factorizable, this time the reduced Wigner distribution function
for $X$ is going to depend on parameters associated with $Y$:
\begin{eqnarray}
 \tilde{\rho}_W (x,p_x) \!\!\!\! & = & \!\!\!\! \frac{A^2}{2\pi\hbar} \
  \bigg\{ e^{-(x-x_1)^2/2\sigma^2 - 2\sigma^2(p_x - p_1^x)^2/\hbar^2}
  + e^{-(x-x_2)^2/2\sigma^2 - 2\sigma^2(p_x - p_2^x)^2/\hbar^2}
  \nonumber \\ \!\!\!\! & & \!\!\!\! \quad \qquad
  + 2 e^{-(\Delta y)^2/8\sigma^2 - \sigma^2(\Delta p_y)^2/2\hbar^2}
  e^{-(x-\bar{x})^2/2\sigma^2 - 2\sigma^2(p_x - \bar{p})^2/\hbar^2}
  \nonumber \\ \!\!\!\! & & \!\!\!\! \qquad \qquad
  \times \cos \left[\frac{ \Delta p_y\!\ \bar{y} + \Delta p_x\!\ x
    - (p_x-\bar{p}_x) \Delta x}{\hbar}\right] \bigg\} .
 \label{eq41}
\end{eqnarray}
Note that this explicit dependence on $Y$ in the interference term implies that two separate phase-space
points in the $Y$-subspace are enough to cancel out this term, eventually leading to the classical
distribution considered for the cat state, Equation~(\ref{eq34}).
Actually, if we choose as the classical Wigner distribution function here the one given by
(\ref{eq40}), but without the crossed interference (coherence) term, i.e.,
\begin{eqnarray}
 \rho_{\rm cl} (x,y,p_x,p_y) \!\!\!\! & = & \!\!\!\! \frac{1}{2} \left(\frac{1}{\pi\hbar}\right)^2
 \left[ e^{-(x-x_1)^2/2\sigma^2 - 2\sigma^2(p_x - p_{x,1}^x)^2/\hbar^2}
        e^{-(y-y_1)^2/2\sigma^2 - 2\sigma^2(p_y - p_{y,1}^y)^2/\hbar^2}
 \right.
 \nonumber \\  \!\!\!\! & & \!\!\!\! \left.
     +  e^{-(x-x_2)^2/2\sigma^2 - 2\sigma^2(p_x - p_{x,2}^x)^2/\hbar^2}
        e^{-(y-y_2)^2/2\sigma^2 - 2\sigma^2(p_y - p_{y,2}^y)^2/\hbar^2}
 \right] ,
 \label{eq42}
\end{eqnarray}
the reduced Wigner distribution is exactly the same as (\ref{eq34}):
\begin{equation}
 \tilde{\rho}_{\rm cl} (x,p_x) = \frac{1}{2\pi\hbar} \
  \bigg\{ e^{-(x-x_1)^2/2\sigma^2 - 2\sigma^2(p_x - p_1)^2/\hbar^2}
  + e^{-(x-x_2)^2/2\sigma^2 - 2\sigma^2(p_x - p_2)^2/\hbar^2} \bigg\} .
 \label{eq43}
\end{equation}
Therefore, in both cases, cat state and entangled state, we eventually reach the same reduced
Wigner distribution function, although the joint Wigner distribution function is different, as
it is a simple delocalization in the $X$-subspace for the former, while it corresponds to a exchange
or swapping in the joint $XY$-space for the latter.

Regarding the initial value of the entropies, now we have
\begin{eqnarray}
 S_L \!\!\!\! & = & \!\!\!\! 1 - \frac{A^4}{2} \
  \bigg[ g_x^2 + g_y^2 - g_x^2 g_y^2 - 1
 + 2 \Big( 1 + g_x g_y \cos \phi {xy} \Big)^2 \bigg] ,
 \label{eq44a} \\
  S_L^{\rm cl} \!\!\!\! & = & \!\!\!\!\frac{1}{2} \left( 1 - g_x^2 \right) ,
 \label{eq44b}
\end{eqnarray}
with
\begin{eqnarray}
 g_x \!\!\!\! & = & \!\!\!\! e^{-(\Delta x)^2/8\sigma^2 - \sigma^2(\Delta p_x)^2/2\hbar^2} ,
  \nonumber \\
 g_y \!\!\!\! & = & \!\!\!\! e^{-(\Delta y)^2/8\sigma^2 - \sigma^2(\Delta p_y)^2/2\hbar^2} ,
  \nonumber \\
 \phi_{xy} \!\!\!\! & = & \!\!\!\! \left( \Delta p_y \bar{y} + \Delta p_x \bar{x} \right)/\hbar .
  \nonumber
\end{eqnarray}
In the case that the phase-space points are far apart, the two values approach the same lower bound,
$S_L = S_L^{\rm cl} = 0.5$.
However, we also find that, even if this separation condition only satisfies for the $Y$ phase-space
points, both entropies coincide, although to the lower bound
\be
 S_L = S_L^{\rm cl} = \frac{1}{2} \left( 1 - g_x^2 \right) ,
\ee
which stresses the fact that decoherence due to the entanglement with $Y$ leads to a delocalized
classical-type state.
Indeed, it is also observed that, as the phase-space points around which the Gaussian distributions are
centered draw closer, and hence the delocalization becomes smaller, this lower bound for the linear entropy
gets smaller.
In the limit of full localization (both Gaussian distributions centered around the same phase-space point),
both linear entropies vanish, which corresponds to the case analyzed in Section~\ref{sec3-2}.
On the other hand, with respect to the von Neumann entropy, in this case neither the quantum entropy
nor the classical one are zero.
The value for the classical one was given above, $S_V^{\rm cl} = 1$, while the quantum value can
readily be obtained by directly applying the second equality of Equation~(\ref{eq2}) to this case, which
renders the well-known result for maximally entangled bipartite systems, $S_V = \ln 2 \approx 0.693$.

\begin{figure}[t]
 \centering
 \includegraphics[width=0.66\textwidth]{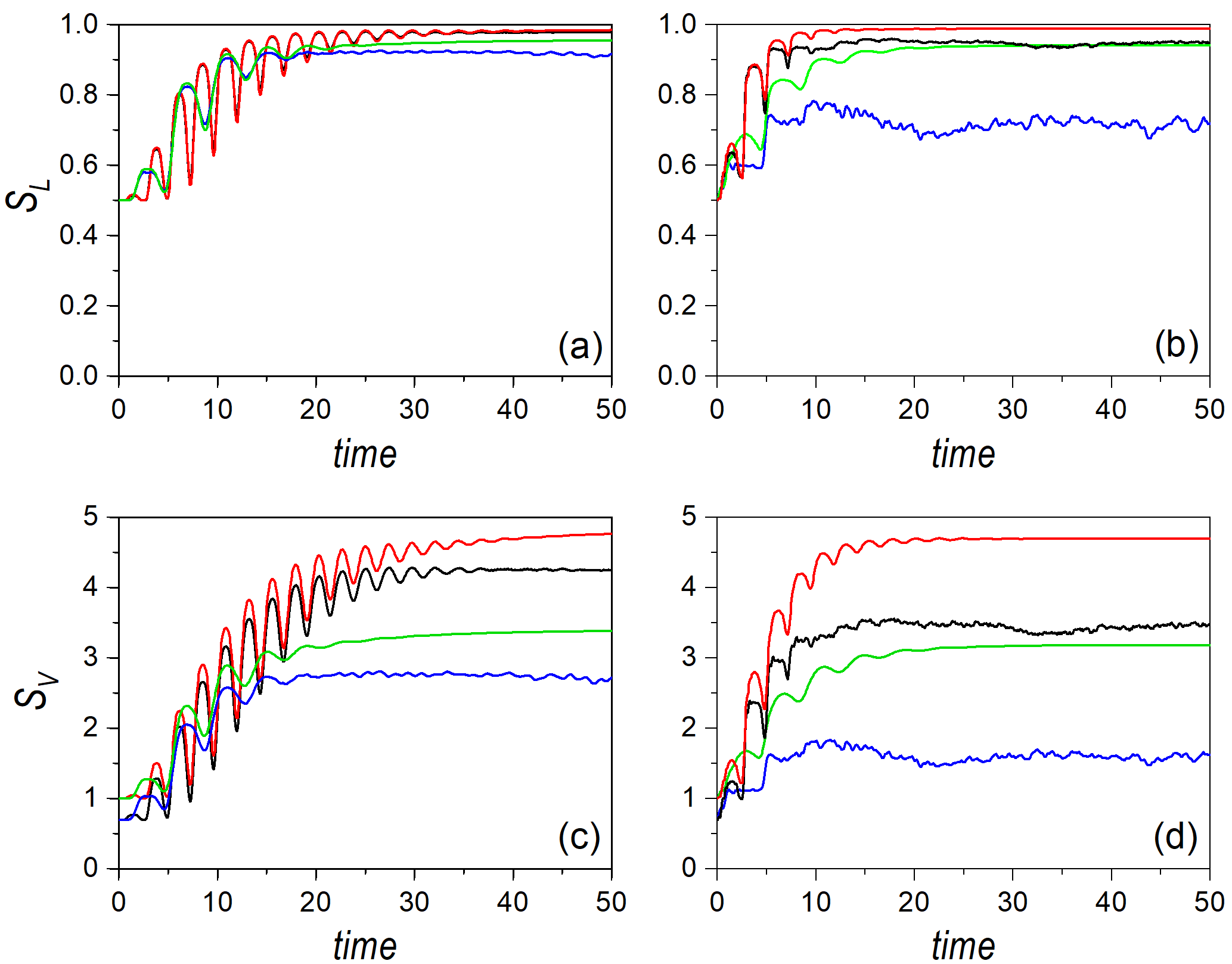}
 \caption{\label{fig7}
 Linear entropy (upper row) and von Neumann entropy (lower row) for an initial
 distribution associated with a Bell-type entangled state for two different values of
 the energy, and affected by regularity conditions (\textbf{a},\textbf{c}) and chaos conditions (\textbf{b},\textbf{d}).
 Results for $E_0=15$ are denoted with the blue and green lines, respectively, for the quantum and
 the classical dynamical regimes.
 Results for $E_0=150$ are denoted with the black and red lines, respectively, for the quantum and
 the classical dynamical regimes.}
\end{figure}

Let us now analyze the behavior displayed by the corresponding phase-space distributions.
To this end, the maximally entangled state selected is specified by positioning the Gaussians at
the phase-space points ${\bf p}_1 = (x_0,0,p_{x,0},0)$ and ${\bf p}_2 = (0,y_0,0,p_{y,0})$, where
$x_0 = y_0 = 2.5$ and $p_{x,0} = p_{y,0} = -\sqrt{2mE'_0}$, with $E'_0 = E_0 - \beta x_0^4/4$.
In this case, in order to better understand the approach to the classical limit by increasing the
total energy, we have considered $E_0=15$ and $E_0=150$.
Note that the initial configuration denotes symmetry with respect to the $y=x$ axis, while the initial
motion of the wave packets takes place along the $x$- and $y$-channels, respectively, which will render
some similarities to former results already discussed.
The time evolution for the quantum and classical entropies corresponding to a maximally entangled
state are displayed in Figure~\ref{fig7}, the linear entropies in the upper panels and the von Neumann
ones in the lower panels; in the left column for the regular regime, and in the right column for the
chaotic one.
In both cases, contrary what one might expect, as we also observed with the cat state, the behavior
here, for both energies, is pretty similar to the behavior found for analogous conditions (i.e., initial
in-channel motion) in the case of a single Gaussian distribution.
Of course, due to the seemingly incoherent addition of the two quantum Gaussian distributions (led by
the decoherence induced by the entanglement between the two subsystems), the initial values are different
compared to those of the initial localized Gaussian, and in compliance with the values analytically
determined and discussed above.
However, as time proceeds, both the oscillatory initial increase and the final asymptotic saturation
regime are pretty similar to the results previously found for the Gaussian distributions, even regarding
the separation between quantum and classical counterparts.

\section{Discussion}
\label{sec4}

As explained in the introductory section, the main goal of the present work was to explore
whether there is any sort of quantum--classical correspondence in the case of the entropy measures
commonly used to quantify entanglement in (joint) pure bipartite systems, namely, the linear and the
von Neumann entropies.
As entanglement is a unique quantum feature, in principle, one should assume that such correspondence
does not take place.
On the other hand, we also know that, as decoherence effects become stronger due to the entanglement
between the two parties, the manifestation of quantum--classical correspondence should be more
apparent.
Thus, with the purpose to compare on equal footing, and hence to determine whether any quantum--classical
correspondence can be established, two coupled continuous-variable bipartite systems have been considered,
consisting of two one-dimensional quartic oscillators nonlinearly coupled by means of a biquadratic term,
which has been shown to display either regular or chaotic dynamics depending on the coupling strength.
The entropy measures have then been recast in terms phase-space distributions, which are a common ground
for comparison purposes to both quantum and classical mechanics, even though there are still certain
differences and subtleties associated with the coherence properties of quantum systems (e.g., the
negativity of the Wigner distribution function, which is lacking in classical phase-space density
distributions).
Furthermore, it has also been necessary to define classical analogs of the initial Wigner distribution
functions, which has been achieved by simply removing the coherent part in the latter, i.e., suppressing
the partly negative-definite interference terms.

With those ingredients, the time evolution displayed by the linear and von Neumann entropies associated
with three particular types of initial states have been analyzed:
\begin{itemize}
 \item Gaussian states, which have been chosen, because these factorizable, localized states keep a very
 close connection to classical distributions of the same kind, thus allowing us to better understand how
 the development and presence of interference is at the origin of the long term differences between the
 quantum and classical asymptotic entropy values.
 In the short and medium terms, however, both the quantum and classical results are mainly governed by
 the oscillations of the system inside the potential in a similar manner.
 Nonetheless, in all cases, the classical values constitute, at each time, an upper bound for the
 quantum ones, which draw closer to the former in the case of the linear entropy, but present a gap
 in the case of the von Neumann entropy.
 This gap becomes smaller as the total energy becomes larger (the system becomes more classical), although
 there is not a clear trend denoting that it vanishes, as in the case of the linear entropy.
 Hence, the gap can be associated with the coherence characterizing the quantum system, which allows it
 to reduce its entropy unlike the classical system, where its incoherence does not allow lower values of
 the entropy.
 This is something that we also observe in the initial values of the von Neumann entropy, which is zero
 in the quantum case (minimum amount of information to specify the state) and nonzero for the classical.
 Note that, although the initial state is the same in both cases (an everywhere positive-definite
 Gaussian distribution), the appearance of the third term in the last equality of (\ref{demo}), which
 is responsible for interference-related contributions (hence leading to the inequality
 $\left( \ln \tilde{\rho} \right)_W \ne \ln \tilde{\rho}_W$), sets a major difference.
 This is, nonetheless, different from the appearance of the HOC term in (\ref{eq1}), which is
 responsible for the development of interference-type features in relation to the interaction potential
 (otherwise, the dynamical evolution would be identical).

 \item Cat states (for the reference system), which has been selected because, regardless of factorizability, they constitute a first approach to delocalization with presence of coherence.
 Although the quantum system is delocalized, because the two (quantum) Gaussian amplitudes are coherently
 superimposed, there is a share of mutual phase information, which translates into a common intermediate
 region in phase space that is partly negative definite.
 The classical counterpart chosen is the truncated Wigner distribution function that arises when such
 interference term is removed.
 Interestingly, similar trends to those observed with the Gaussian distributions are also present here,
 which clearly indicate some degree of correspondence.
 However, because the classical distribution is delocalized, none of the entropy measures renders a
 vanishing value, as it does in the quantum case.
 This thus supports the fact that the entropy measures, apart from dynamical information (e.g., the
 oscillatory behavior in the short and medium term, and the saturation due to ergodicity in the long
 term), also contain information about the coherence of the distribution and the degree of
 localization.
 Note, in this latter regard, that the initial value for the classical linear entropy does not vanish,
 as it does for single Gaussian distribution.

 \item Bell-type entangled states, which constitute a paradigm of maximally entangled states, and that,
 as it has been pointed out, generate delocalization of the reference system with suppression of the
 interference term if the two possible phase-space locations for the environmental degree of freedom
 are sufficiently far from each other.
 In this case, the classical distribution, which is the same ansatz also associated with the cat state,
 looks exactly the same as the quantum Wigner distribution function.
 However, unlike the case of the cat state, now the suppression of the interference term by entanglement
 provides us with a more ``classical'' state, as the two density distributions that appear in phase
 space are incoherent (coherence is only preserved in the joint space).
 Yet, although we are talking about a maximally entangled state, and both the linear and the von Neumann
 entropies behave as expected quantum mechanically, rendering nonzero initial values, again we find a
 good correspondence with the classical counterparts in all cases considered.
 In fact, leaving aside the initial values, the trend observed in time again matches the
 trends obtained for the corresponding single Gaussian distributions rather well.
\end{itemize}


\section{Concluding Remarks}
\label{sec5}

In summary, in the light of the above results, we can say that there is not a major difference between
the quantum and classical results in a situation where, in principle, one would assume that there is no
classical analog. In fact, we conclude just the opposite, that it seems there is a nice
quantum--classical correspondence.
Therefore, the widespread idea that the linear and von Neumann entropies are good measures of entanglement
seems to fail, as they can be fairly reproduced classically by means of simple classical analogs,
obtained from the suppression of the coherence (interference) terms in the corresponding Wigner
distribution functions.
Indeed, at any time, it has been seen that the classical value is always above the quantum mechanical one,
which could be wrongly interpreted, according to our current understanding of these entropies, as a more
efficient production of entanglement from classical states than from quantum ones.
After all, it has been common to regard classically chaotic regimes as more efficient in the production of
entanglement than regular ones by inspecting the same kind of entropies and their time evolution (although
here it has been proven that not always this is true, for there is a crossover that depends on the total
energy available to the joint system).
Therefore, from the results here obtained, particular confronting the simulations shown in Section~\ref{sec3-4}
with those from Section~\ref{sec3-2}, it seems that these entropy measures, actually, quantify the degrees
of delocalization and incoherence that affect the system, which are present in both the quantum and the
classical domains.
Of course, in the quantum domain, because the full joint system preserves its coherence, the interaction
increases the correlations between both and, hence, the entanglement, which is nothing but a coherence
swapping or redistribution between both subsystems, in such a way that its effective manifestation in any
of them is in the form of decoherence.
In the subsystem of reference, decoherence washes out interference-related terms, thus producing a
seemingly classical distribution.
Yet, the preservation of coherence at a higher order (i.e.., in the phase space of the complete system)
makes a difference in entropy measures that emphasize such orders, such as the von Neumann entropy, which
thus remains manifestly lower than its classical partner.

\vspace{6pt}
\noindent
{\bf Funding:}
`This research was funded by the European Union’s Horizon 2020 Research and Innovation
Program under the QuantERA programme, through the project ApresSF, and from the EU Grant 899587 (Project
Stormytune) and the Spanish Agencia Estatal de Investigaci\'on (Grant No.\ PCI2019-111874-2).


\conflictsofinterest{The author declares no conflict of interest.}


\reftitle{References}


\begin{thebibliography}{999}
\providecommand{\natexlab}[1]{#1}

\bibitem[Schr\"odinger(1935)]{schrodinger:ProcCamPS:1935}
Schr\"odinger, E.
\newblock Discussion of probability relations between separated systems.
\newblock {\em Math. Proc. Camb. Philos. Soc.} {\bf 1935}, {\em
  31},~555--563.
  doi:{\changeurlcolor{black}\href{https://doi.org/10.1017/S0305004100013554}{\detokenize{10.1017/S0305004100013554}}}.

\bibitem[Bouwmeester \em{et~al.}(2000)Bouwmeester, Ekert, and
  Zeilinger]{bouwmeester-bk:2000}
Bouwmeester, D.; Ekert, A.; Zeilinger, A. (Eds.)
\newblock {\em The Physics of Quantum Information}; Springer: Berlin, Germany, 2000.

\bibitem[Nielsen and Chuang(2000)]{nielsen-chuang-bk}
Nielsen, M.A.; Chuang, I.L.
\newblock {\em Quantum Computation and Quantum Information}; Cambridge
  University Press: Cambridge, UK, 2000.

\bibitem[Einstein \em{et~al.}(1935)Einstein, Podolsky, and
  Rosen]{EPR:PhysRev:1935}
Einstein, A.; Podolsky, B.; Rosen, N.
\newblock Can quantum-mechanical description of physical reality be considered
  complete?
\newblock {\em Phys. Rev.} {\bf 1935}, {\em 47},~777--780.
  doi:{\changeurlcolor{black}\href{https://doi.org/10.1103/PhysRev.47.777}{\detokenize{10.1103/PhysRev.47.777}}}.

\bibitem[Zurek and Paz(1995)]{zurek:PhysicaD:1995}
Zurek, W.H.; Paz, J.P.
\newblock Quantum chaos: A decoherent definition.
\newblock {\em Phys. D Nonlinear Phenom.} {\bf 1995}, {\em 83},~300--308.
  doi:{\changeurlcolor{black}\href{https://doi.org/10.1016/0167-2789(94)00271-Q}{\detokenize{10.1016/0167-2789(94)00271-Q}}}.

\bibitem[Furuya \em{et~al.}(1998)Furuya, Nemes, and
  Pellegrino]{furuya:PRL:1998}
Furuya, K.; Nemes, M.C.; Pellegrino, G.Q.
\newblock Quantum Dynamical Manifestation of Chaotic Behavior in the Process of
  Entanglement.
\newblock {\em Phys. Rev. Lett.} {\bf 1998}, {\em 80},~5524--5527.
\newblock
  doi:{\changeurlcolor{black}\href{https://doi.org/10.1103/PhysRevLett.80.5524}{\detokenize{10.1103/PhysRevLett.80.5524}}}.

\bibitem[Miller and Sarkar(1999)]{sarkar:PRE:1999}
Miller, P.A.; Sarkar, S.
\newblock Signatures of chaos in the entanglement of two coupled quantum kicked
  tops.
\newblock {\em Phys. Rev. E} {\bf 1999}, {\em 60},~1542--1550.
\newblock
  doi:{\changeurlcolor{black}\href{https://doi.org/10.1103/PhysRevE.60.1542}{\detokenize{10.1103/PhysRevE.60.1542}}}.

\bibitem[Ghose and Sanders(2004)]{ghose:PRA:2004}
\textls[-25]{Ghose, S.; Sanders, B.C.
\newblock Entanglement dynamics in chaotic systems.
\newblock {\em Phys. Rev. A} {\bf 2004}, {\em 70},~062315.
\newblock
  doi:{\changeurlcolor{black}\href{https://doi.org/10.1103/PhysRevA.70.062315}{\detokenize{10.1103/PhysRevA.70.062315}}}.}

\bibitem{ghose:PRE:2004}
Wang, X.; Ghose, S.; Sanders, B.C.; Hu, B.
\newblock Entanglement as a signature of quantum chaos.
\newblock {\em Phys. Rev. E} {\bf 2004}, {\em 70},~016217.
\newblock
  doi:{\changeurlcolor{black}\href{https://doi.org/10.1103/PhysRevE.70.016217}{\detokenize{10.1103/PhysRevE.70.016217}}}.

\bibitem[Weyl(1927)]{weyl:ZPhys:1927}
Weyl, H.
\newblock Quantenmechanik und Gruppentheorie.
\newblock {\em Z. Phys.} {\bf 1927}, {\em 46},~1--46.
\newblock
  doi:{\changeurlcolor{black}\href{https://doi.org/10.1007/BF02055756}{\detokenize{10.1007/BF02055756}}}.

\bibitem[Wigner(1932)]{wigner:PhysRev:1932}
\textls[-15]{Wigner, E.
\newblock On the Quantum Correction For Thermodynamic Equilibrium.
\newblock {\em Phys. Rev.} {\bf 1932}, {\em 40},~749--759.
\newblock
  doi:{\changeurlcolor{black}\href{https://doi.org/10.1103/PhysRev.40.749}{\detokenize{10.1103/PhysRev.40.749}}}.}

\bibitem[Moyal(1949)]{moyal:MathProcCambrdPhilSoc:1949}
\textls[-15]{Moyal, J.E.
\newblock Quantum mechanics as a statistical theory.
\newblock {\em Math. Proc. Camb. Philos. Soc.}
  {\bf 1949}, {\em 45},~99--124.
\newblock
  doi:{\changeurlcolor{black}\href{https://doi.org/10.1017/S0305004100000487}{\detokenize{10.1017/S0305004100000487}}}.}

\bibitem[Ballentine(1998)]{ballentine-bk}
Ballentine, L.E.
\newblock {\em Quantum Mechanics. A Modern Development}; World Scientific:
  Singapore,  1998.

\bibitem[Jaffe and Brumer(1984)]{brumer-jaffe:JPC:1984}
Jaff\'e, C.; Brumer, P.
\newblock Classical Liouville mechanics and intramolecular relaxation dynamics.
\newblock {\em J. Phys. Chem.} {\bf 1984}, {\em 88},~4829--4839.
\newblock
  doi:{\changeurlcolor{black}\href{https://doi.org/10.1021/j150665a007}{\detokenize{10.1021/j150665a007}}}.

\bibitem[Jaff\'e and Brumer(1985)]{brumer-jaffe:JCP:1985}
Jaff\'e, C.; Brumer, P.
\newblock Classical-quantum correspondence in the distribution dynamics of
  integrable systems.
\newblock {\em J. Chem. Phys.} {\bf 1985}, {\em 82},~2330--2340.
\newblock
  doi:{\changeurlcolor{black}\href{https://doi.org/10.1063/1.448946}{\detokenize{10.1063/1.448946}}}.

\bibitem[Jaff\'e \em{et~al.}(1985)Jaff\'e, Kanfer, and
  Brumer]{brumer-jaffe:PRL:1985}
Jaff\'e, C.; Kanfer, S.; Brumer, P.
\newblock Classical Analog of Pure-State Quantum Dynamics.
\newblock {\em Phys. Rev. Lett.} {\bf 1985}, {\em 54},~8--10.
\newblock
  doi:{\changeurlcolor{black}\href{https://doi.org/10.1103/PhysRevLett.54.8}{\detokenize{10.1103/PhysRevLett.54.8}}}.

\bibitem[Milburn(1986)]{milburn:PRA:1986}
Milburn, G.J.
\newblock Quantum and classical Liouville dynamics of the anharmonic
  oscillator.
\newblock {\em Phys. Rev. A} {\bf 1986}, {\em 33},~674--685.
\newblock
  doi:{\changeurlcolor{black}\href{https://doi.org/10.1103/PhysRevA.33.674}{\detokenize{10.1103/PhysRevA.33.674}}}.

\bibitem[Gong and Brumer(2003{\natexlab{a}})]{brumer-gong:PRL:2003}
Gong, J.; Brumer, P.
\newblock When is Quantum Decoherence Dynamics Classical?
\newblock {\em Phys. Rev. Lett.} {\bf 2003}, {\em 90},~050402.
\newblock
  doi:10.1103/PhysRev\\Lett.90.050402.

\bibitem[Gong and Brumer(2003{\natexlab{b}})]{brumer-gong:PRA:2003}
Gong, J.; Brumer, P.
\newblock Intrinsic decoherence dynamics in smooth Hamiltonian systems:
  Quantum-classical correspondence.
\newblock {\em Phys. Rev. A} {\bf 2003}, {\em 68},~022101.
\newblock
  doi:{\changeurlcolor{black}\href{https://doi.org/10.1103/PhysRevA.68.022101}{\detokenize{10.1103/PhysRevA.68.022101}}}.

\bibitem[Gong and Brumer(2003{\natexlab{c}})]{brumer-gong:JModOpt:2003}
\textls[-25]{Gong, J.; Brumer, P.
\newblock Quantum versus classical decoherence dynamics.
\newblock {\em J. Mod. Opt.} {\bf 2003}, {\em 50},~2411--2422.
\newblock
  doi:{\changeurlcolor{black}\href{https://doi.org/10.1080/09500340308233572}{\detokenize{10.1080/09500340308233572}}}.}

\bibitem[Hamilton and Brumer(1982)]{brumer-hamilton:PRA:1982}
Hamilton, I.; Brumer, P.
\newblock Relaxation rates for two-dimensional deterministic mappings.
\newblock {\em Phys. Rev. A} {\bf 1982}, {\em 25},~3457--3459.
\newblock
  doi:{\changeurlcolor{black}\href{https://doi.org/10.1103/PhysRevA.25.3457}{\detokenize{10.1103/PhysRevA.25.3457}}}.

\bibitem[Hamilton and Brumer(1983)]{brumer-hamilton:JCP:1983}
Hamilton, I.; Brumer, P.
\newblock Intramolecular relaxation in N = 2 Hamiltonian systems: The role of the
  K entropy.
\newblock {\em J. Chem. Phys.} {\bf 1983}, {\em
  78},~2682--2690.
\newblock
  doi:{\changeurlcolor{black}\href{https://doi.org/10.1063/1.445027}{\detokenize{10.1063/1.445027}}}.

\bibitem[Christoffel and Brumer(1986)]{brumer-christoffel:PRA:1986}
Christoffel, K.M.; Brumer, P.
\newblock Quantum and classical dynamics in the stadium billiard.
\newblock {\em Phys. Rev. A} {\bf 1986}, {\em 33},~1309--1321.
\newblock
  doi:{\changeurlcolor{black}\href{https://doi.org/10.1103/PhysRevA.33.1309}{\detokenize{10.1103/PhysRevA.33.1309}}}.

\bibitem[Pattanayak and Brumer(1996)]{brumer-pattnayak:PRL:1996}
Pattanayak, A.K.; Brumer, P.
\newblock Exponential Divergence and Long Time Relaxation in Chaotic Quantum
  Dynamics.
\newblock {\em Phys. Rev. Lett.} {\bf 1996}, {\em 77},~59--62.
\newblock
  doi:{\changeurlcolor{black}\href{https://doi.org/10.1103/PhysRevLett.77.59}{\detokenize{10.1103/PhysRevLett.77.59}}}.

\bibitem[Pattanayak and Brumer(1997{\natexlab{a}})]{brumer-pattnayak:PRL:1997}
Pattanayak, A.K.; Brumer, P.
\newblock Exponentially Rapid Decoherence of Quantum Chaotic Systems.
\newblock {\em Phys. Rev. Lett.} {\bf 1997}, {\em 79},~4131--4134.
\newblock
  doi:{\changeurlcolor{black}\href{https://doi.org/10.1103/PhysRevLett.79.4131}{\detokenize{10.1103/PhysRevLett.79.4131}}}.

\bibitem[Pattanayak and Brumer(1997{\natexlab{b}})]{brumer-pattnayak:PRE:1997}
Pattanayak, A.K.; Brumer, P.
\newblock Chaos and Lyapunov exponents in classical and quantal distribution
  dynamics.
\newblock {\em Phys. Rev. E} {\bf 1997}, {\em 56},~5174--5177.
\newblock
  doi:{\changeurlcolor{black}\href{https://doi.org/10.1103/PhysRevE.56.5174}{\detokenize{10.1103/PhysRevE.56.5174}}}.

\bibitem[Bettelli and Shepelyansky(2003)]{shepelyansky:PRA:2003}
Bettelli, S.; Shepelyansky, D.L.
\newblock Entanglement versus relaxation and decoherence in a quantum algorithm
  for quantum chaos.
\newblock {\em Phys. Rev. A} {\bf 2003}, {\em 67},~054303.
\newblock
  doi:{\changeurlcolor{black}\href{https://doi.org/10.1103/PhysRevA.67.054303}{\detokenize{10.1103/PhysRevA.67.054303}}}.

\bibitem[Angelo and Furuya(2005)]{furuya:PRA:2005}
Angelo, R.M.; Furuya, K.
\newblock Semiclassical limit of the entanglement in closed pure systems.
\newblock {\em Phys. Rev. A} {\bf 2005}, {\em 71},~042321.
\newblock
  doi:{\changeurlcolor{black}\href{https://doi.org/10.1103/PhysRevA.71.042321}{\detokenize{10.1103/PhysRevA.71.042321}}}.

\bibitem{dodonov:Physica:1974}
 Dodonov, V.V.; Malkin, I.A.; Man'ko, V.I.
\newblock Even and odd coherent states and excitations of a singular oscillator.
\newblock {\em Physica} {\bf 1974}, {\em 72},~597--615.
\newblock
  doi:{\changeurlcolor{black}\href{http://dx.doi.org/10.1016/0031-8914(74)90215-8}{\detokenize{10.1016/0031-8914(74)90215-8}}}.

\bibitem{dodonov-manko-bk:2003}
 Dodonov, V.V.; Man'ko, V.I. (Eds.)
\newblock \emph{Theory of Nonclassical States of Light};
 Taylor \& Francis: London, UK; New York, NY, USA, 2003.

\bibitem{sanders:PRA:1992}
 Sanders, B.C.
\newblock Entangled coherent states.
\newblock {\em Phys. Rev. A} {\bf 1992}, {\em 45},~6811--6815.
\newblock
  doi:{\changeurlcolor{black}\href{http://dx.doi.org/10.1103/PhysRevA.45.6811}{\detokenize{10.1103/PhysRevA.45.6811}}};
\newblock Erratum in {\em Phys. Rev. A} {\bf 1992}, {\em 46},~2966.
\newblock
  doi:{\changeurlcolor{black}\href{http://dx.doi.org/10.1103/PhysRevA.46.2966}{\detokenize{10.1103/PhysRevA.46.2966}}}.

\bibitem{sanders:JMO:1993}
 Wielinga, B.; Sanders, B.C.
\newblock Entangled coherent states with variable weighting.
\newblock {\em J. Mod. Opt.} {\bf 1993}, {\em 40},~1923--1937.
\newblock
  doi:{\changeurlcolor{black}\href{http://dx.doi.org/10.1080/09500349314551951}{\detokenize{10.1080/09500349314551951}}}.

\bibitem{sanders:JPhysA:2012}
 Sanders, B.C.
\newblock Review of entangled coherent states.
\newblock {\em J. Phys. A Math. Theor.} {\bf 2012}, {\em 45},~244002.
\newblock
  doi:{\changeurlcolor{black}\href{http://dx.doi.org/10.1088/1751-8113/45/24/244002}{\detokenize{10.1088/1751-8113/45/24/244002}}}.

\bibitem[Vedral \em{et~al.}(1997)Vedral, Plenio, Rippin, and
  Knight]{plenio-vedral:PRL:1997}
Vedral, V.; Plenio, M.B.; Rippin, M.A.; Knight, P.L.
\newblock Quantifying entanglement.
\newblock {\em Phys. Rev. Lett.} {\bf 1997}, {\em 78},~2275--2279.
\newblock
  doi:{\changeurlcolor{black}\href{https://doi.org/10.1103/PhysRevLett.78.2275}{\detokenize{10.1103/PhysRevLett.78.2275}}}.

\bibitem[Vedral and Plenio(1998)]{plenio-vedral:PRA:1998}
Vedral, V.; Plenio, M.B.
\newblock Entanglement measures and purification procedures.
\newblock {\em Phys. Rev. A} {\bf 1998}, {\em 57},~1619--1633.
\newblock
  doi:{\changeurlcolor{black}\href{https://doi.org/10.1103/PhysRevA.57.1619}{\detokenize{10.1103/PhysRevA.57.1619}}}.

\bibitem[Horodecki(2001)]{horodecki:QuantumInfComp:2001}
Horodecki, M.
\newblock Entanglement measures.
\newblock {\em Quantum Inf. Comput.} {\bf 2001}, {\em 1},~3--26.
\newblock
  doi:{\changeurlcolor{black}\href{https://doi.org/10.26421/QIC1.1-2}{\detokenize{10.26421/QIC1.1-2}}}.

\bibitem[Bru{\ss}(2002)]{bruss:JMathPhys:2002}
Bru{\ss}, D.
\newblock Characterizing entanglement.
\newblock {\em J. Math. Phys.} {\bf 2002}, {\em
  43},~4237--4251.
\newblock
  doi:{\changeurlcolor{black}\href{https://doi.org/10.1063/1.1494474}{\detokenize{10.1063/1.1494474}}}.

\bibitem[Wehrl(1978)]{wehrl:RMP:1978}
Wehrl, A.
\newblock General properties of entropy.
\newblock {\em Rev. Mod. Phys.} {\bf 1978}, {\em 50},~221--260.
\newblock
  doi:{\changeurlcolor{black}\href{https://doi.org/10.1103/RevModPhys.50.221}{\detokenize{10.1103/RevModPhys.50.221}}}.

\bibitem[Wehrl(1979)]{wehrl:RepMathPhys:1979}
Wehrl, A.
\newblock On the relation between classical and quantum-mechanical entropy.
\newblock {\em Rep. Math. Phys.} {\bf 1979}, {\em 16},~353--358.
\newblock
  doi:{\changeurlcolor{black}\href{https://doi.org/10.1016/0034-4877(79)90070-3}{\detokenize{10.1016/0034-4877(79)90070-3}}}.

\bibitem[Wehrl(1991)]{wehrl:RepMathPhys:1991}
Wehrl, A.
\newblock The many facets of entropy.
\newblock {\em Rep. Math. Phys.} {\bf 1991}, {\em 30},~119--129.
\newblock
  doi:{\changeurlcolor{black}\href{https://doi.org/10.1016/0034-4877(91)90045-O}{\detokenize{10.1016/0034-4877(91)90045-O}}}.

\bibitem[Popescu and Rohrlich(1997)]{popescu-rohrlich:PRA:1997}
Popescu, S.; Rohrlich, D.
\newblock Thermodynamics and the measure of entanglement.
\newblock {\em Phys. Rev. A} {\bf 1997}, {\em 56},~R3319--R3321.
\newblock
  doi:{\changeurlcolor{black}\href{https://doi.org/10.1103/PhysRevA.56.R3319}{\detokenize{10.1103/PhysRevA.56.R3319}}}.

\bibitem[Bennett \em{et~al.}(1996{\natexlab{a}})Bennett, DiVincenzo, Smolin,
  and Wootters]{bennett-smolin:PRA:1996}
Bennett, C.H.; DiVincenzo, D.P.; Smolin, J.A.; Wootters, W.K.
\newblock Mixed-state entanglement and quantum error correction.
\newblock {\em Phys. Rev. A} {\bf 1996}, {\em 54},~3824--3851.
\newblock
  doi:{\changeurlcolor{black}\href{https://doi.org/10.1103/PhysRevA.54.3824}{\detokenize{10.1103/PhysRevA.54.3824}}}.

\bibitem{bennett-smolin:PRL:1996}
 Bennett, C.H.; Brassard, G.; Popescu, S.; Schumacher, B.; Smolin, J.A.; Wootters, W.K.
\newblock Purification of Noisy Entanglement and Faithful Teleportation via Noisy Channels.
\newblock {\em Phys. Rev. Lett.} {\bf 1996}, {\em 76},~722--725.
\newblock doi:{\changeurlcolor{black}\href{https://doi.org/10.1103/PhysRevLett.76.722}{\detokenize{10.1103/PhysRevLett.76.722}}}.
\newblock Erratum in {\em Phys. Rev. Lett.} {\bf 1997}, {\em 78},~2031.
\newblock
  doi:{\changeurlcolor{black}\href{https://doi.org/10.1103/PhysRevLett.78.2031}{\detokenize{10.1103/PhysRevLett.78.2031}}}.




\bibitem[Wei \em{et~al.}(2003)Wei, Nemoto, Goldbart, Kwiat, Munro, and
  Verstraete]{verstraete:PRA:2003}
Wei, T.C.; Nemoto, K.; Goldbart, P.M.; Kwiat, P.G.; Munro, W.J.; Verstraete, F.
\newblock Maximal entanglement versus entropy for mixed quantum states.
\newblock {\em Phys. Rev. A} {\bf 2003}, {\em 67},~022110.
\newblock
  doi:{\changeurlcolor{black}\href{https://doi.org/10.1103/PhysRevA.67.022110}{\detokenize{10.1103/PhysRevA.67.022110}}}.

\bibitem[Peres(1996)]{peres:PRL:1996}
Peres, A.
\newblock Separability criterion for density matrices.
\newblock {\em Phys. Rev. Lett.} {\bf 1996}, {\em 77},~1413--1415.
\newblock
  doi:{\changeurlcolor{black}\href{https://doi.org/10.1103/PhysRevLett.77.1413}{\detokenize{10.1103/PhysRevLett.77.1413}}}.

\bibitem[Horodecki \em{et~al.}(1996)Horodecki, Horodecki, and
  Horodecki]{horodecki:PhysLettA:1996}
Horodecki, M.; Horodecki, P.; Horodecki, R.
\newblock Separability of mixed states: necessary and sufficient conditions.
\newblock {\em Phys. Lett. A} {\bf 1996}, {\em 223},~1--8.
\newblock
  doi:{\changeurlcolor{black}\href{https://doi.org/10.1016/S0375-9601(96)00706-2}{\detokenize{10.1016/S0375-9601(96)00706-2}}}.

\bibitem[Duan \em{et~al.}(2000)Duan, Giedke, Cirac, and
  Zoller]{duan-cirac-zoller:PRL:2000}
Duan, L.M.; Giedke, G.; Cirac, J.I.; Zoller, P.
\newblock Inseparability Criterion for Continuous Variable Systems.
\newblock {\em Phys. Rev. Lett.} {\bf 2000}, {\em 84},~2722--2725.
\newblock
  doi:{\changeurlcolor{black}\href{https://doi.org/10.1103/PhysRevLett.84.2722}{\detokenize{10.1103/PhysRevLett.84.2722}}}.

\bibitem[Simon(2000)]{simon:PRL:2000}
Simon, R.
\newblock Peres-Horodecki Separability Criterion for Continuous Variable
  Systems.
\newblock {\em Phys. Rev. Lett.} {\bf 2000}, {\em 84},~2726--2729.
\newblock
  doi:{\changeurlcolor{black}\href{https://doi.org/10.1103/PhysRevLett.84.2726}{\detokenize{10.1103/PhysRevLett.84.2726}}}.

\bibitem[Shannon(1948{\natexlab{a}})]{shannon:BellSystTechJ:1948-1}
Shannon, C.E.
\newblock A mathematical theory of communication.
\newblock {\em  Bell Syst. Tech. J.} {\bf 1948}, {\em
  27},~379--423.
\newblock
  doi:{\changeurlcolor{black}\href{https://doi.org/10.1002/j.1538-7305.1948.tb01338.x}{\detokenize{10.1002/j.1538-7305.1948.tb01338.x}}}.

\bibitem[Shannon(1948{\natexlab{b}})]{shannon:BellSystTechJ:1948-2}
Shannon, C.E.
\newblock A mathematical theory of communication.
\newblock {\em  Bell Syst. Tech. J.} {\bf 1948}, {\em
  27},~623--656.
\newblock
  doi:{\changeurlcolor{black}\href{https://doi.org/10.1002/j.1538-7305.1948.tb00917.x}{\detokenize{10.1002/j.1538-7305.1948.tb00917.x}}}.

\bibitem[Meyer(1986)]{HDmeyer:JCP:1986}
Meyer, H.
\newblock Theory of the Liapunov exponents of Hamiltonian systems and a
  numerical study on the transition from regular to irregular classical motion.
\newblock {\em J. Chem. Phys.} {\bf 1986}, {\em
  84},~3147--3161.
\newblock
  doi:{\changeurlcolor{black}\href{https://doi.org/10.1063/1.450296}{\detokenize{10.1063/1.450296}}}.

\bibitem[Dahlqvist and Russberg(1990)]{dahlqvist:PRL:1990}
Dahlqvist, P.; Russberg, G.
\newblock Existence of stable orbits in the
  ${\mathit{x}}^{2}$${\mathit{y}}^{2}$ potential.
\newblock {\em Phys. Rev. Lett.} {\bf 1990}, {\em 65},~2837--2838.
\newblock
  doi:{\changeurlcolor{black}\href{https://doi.org/10.1103/PhysRevLett.65.2837}{\detokenize{10.1103/PhysRevLett.65.2837}}}.

\bibitem[Bohigas \em{et~al.}(1993)Bohigas, Tomsovic, and
  Ullmo]{bohigas:PhysRep:1993}
Bohigas, O.; Tomsovic, S.; Ullmo, D.
\newblock Manifestations of classical phase space structures in quantum
  mechanics.
\newblock {\em Phys. Rep.} {\bf 1993}, {\em 223},~43--133.
\newblock
  doi:{\changeurlcolor{black}\href{https://doi.org/10.1016/0370-1573(93)90109-Q}{\detokenize{10.1016/0370-1573(93)90109-Q}}}.

\bibitem[Eckhardt \em{et~al.}(1989)Eckhardt, Hose, and
  Pollak]{eckhardt-pollak:PRA:1989}
Eckhardt, B.; Hose, G.; Pollak, E.
\newblock Quantum mechanics of a classically chaotic system: Observations on
  scars, periodic orbits, and vibrational adiabaticity.
\newblock {\em Phys. Rev. A} {\bf 1989}, {\em 39},~3776--3793.
\newblock
  doi:{\changeurlcolor{black}\href{https://doi.org/10.1103/PhysRevA.39.3776}{\detokenize{10.1103/PhysRevA.39.3776}}}.

\bibitem[Joy and Sabir(1993)]{joy:ModPhysLettB:1993}
Joy, M.P.; Sabir, M.
\newblock Chaos and quantum fluctuations in a quartic Hamiltonian system.
\newblock {\em Mod. Phys. Lett. B} {\bf 1993}, {\em 7},~1421--1427.
\newblock
  doi:{\changeurlcolor{black}\href{https://doi.org/10.1142/S0217984993001466}{\detokenize{10.1142/S0217984993001466}}}.

\bibitem[de~Polavieja \em{et~al.}(1994)de~Polavieja, Borondo, and
  Benito]{polavieja-borondo:PRL:1994}
de~Polavieja, G.G.; Borondo, F.; Benito, R.M.
\newblock Scars in Groups of Eigenstates in a Classically Chaotic System.
\newblock {\em Phys. Rev. Lett.} {\bf 1994}, {\em 73},~1613--1616.
\newblock
  doi:{\changeurlcolor{black}\href{https://doi.org/10.1103/PhysRevLett.73.1613}{\detokenize{10.1103/PhysRevLett.73.1613}}}.

\bibitem[Sanz and Miret-Art\'es(2014)]{sanz-bk-2}
Sanz, A.S.; Miret-Art\'es, S.
\newblock {\em A Trajectory Description of Quantum Processes. II.
  Applications}; {Lecture Notes in Physics}; Springer: Berlin,
  Germany, 2014; Volume 831.

\bibitem[Sanz and Miret-Art\'es(2008)]{sanz:JPA:2008}
Sanz, A.S.; Miret-Art\'es, S.
\newblock A trajectory-based understanding of quantum interference.
\newblock {\em J. Phys. A Math. Theor.} {\bf 2008}, {\em 41},~435303.
\newblock
  doi:{\changeurlcolor{black}\href{https://doi.org/10.1088/1751-8113/41/43/435303}{\detokenize{10.1088/1751-8113/41/43/435303}}}.

\bibitem[Sanz \em{et~al.}(2012)Sanz, Elran, and Brumer]{sanz:PRE:2012}
Sanz, A.S.; Elran, Y.; Brumer, P.
\newblock Temperature crossover of decoherence rates in chaotic and regular
  bath dynamics.
\newblock {\em Phys. Rev. E} {\bf 2012}, {\em 85},~036218.
\newblock
  doi:{\changeurlcolor{black}\href{https://doi.org/10.1103/PhysRevE.85.036218}{\detokenize{10.1103/PhysRevE.85.036218}}}.

\end{thebibliography}


\externalbibliography{no}

\end{document}